\documentclass[a4paper,fleqn,usenatbib]{mnras}

\usepackage{newtxtext,newtxmath}

\usepackage[T1]{fontenc}
\usepackage{ae,aecompl}

\usepackage{graphicx}	
\usepackage{amsmath}	
\usepackage{amssymb}	

\usepackage{amsmath}
\usepackage{xcolor}
\usepackage{url}
\usepackage{pdfpages}
\usepackage{graphicx}
\usepackage{booktabs}
\usepackage{enumitem}
\usepackage{array}
\usepackage{makecell}
\usepackage{epstopdf}

\usepackage{dcolumn}
\usepackage{bm}
\usepackage{amsmath}
\usepackage{eqnarray}
\usepackage{array}
\usepackage{makecell}
\usepackage{tabu}
\usepackage{color}

\newcolumntype{L}[1]{>{\raggedright\let\newline\\\arraybackslash\hspace{0pt}}m{#1}}
\newcolumntype{C}[1]{>{\centering\let\newline\\\arraybackslash\hspace{0pt}}m{#1}}
\newcolumntype{R}[1]{>{\raggedleft\let\newline\\\arraybackslash\hspace{0pt}}m{#1}}

\newcommand{\eg}{e.g., }

\newcommand{\ie}{i.e., }

\newcommand{\sect}[1]{Section \ref{s:#1}}

\newcommand{\fig}[1]{Fig.\ \ref{fig:#1}}

\newcommand{\Fig}[1]{Figure \ref{fig:#1}}

\newcommand{\tbl}[1]{Table \ref{t:#1}}

\newcommand{\hide}[1]{} 

\title{An experimental study of low-velocity impacts into granular material in reduced gravity}

\author[N. Murdoch et al.]{
Naomi Murdoch$^{1}$\thanks{E-mail: naomi.murdoch@isae.fr},
Iris Avila Martinez$^{1}$,
Cecily Sunday$^{1,2}$,
Emmanuel Zenou$^{3}$,
\newauthor Olivier Cherrier$^{4}$, 
Alexandre Cadu$^{1}$
and Yves Gourinat$^{4}$\\
\\
\\
$^{1}$D\'{e}partement Electronique, Optronique et Signal (DEOS), Institut Sup\'{e}rieur de l'A\'{e}ronautique et de l'Espace (ISAE-SUPAERO), \\ Universit\'{e} de Toulouse, 31055 Toulouse Cedex 4, France\\
$^{2}$Jet Propulsion Laboratory, 4800 Oak Grove Drive, Pasadena, CA, 91107, USA \\
$^{3}$D\'{e}partement d'Ing\'{e}nierie des Syst\`{e}mes Complexes (DISC), Institut Sup\'{e}rieur de l'A\'{e}ronautique et de l'Espace (ISAE-SUPAERO), \\ Universit\'{e} de Toulouse, 31055 Toulouse Cedex 4, France\\
$^{4}$D\'{e}partement M\'{e}canique des Structures et Mat\'{e}riaux (DMSM), Institut Sup\'{e}rieur de l'A\'{e}ronautique et de l'Espace (ISAE-SUPAERO), \\ Universit\'{e} de Toulouse, 31055 Toulouse Cedex 4, France
}

\date{Accepted 2016 December 30. Published 2017 Jan 4.}

\pubyear{2017}

\begin{document}
\label{firstpage}
\pagerange{\pageref{firstpage}--\pageref{lastpage}}
\maketitle

\begin{abstract}
In order to improve our understanding of landing on small bodies and of asteroid evolution, we use our novel drop tower facility \citep{sunday2016} to perform low-velocity (2 - 40 cm/s), shallow impact experiments of a 10 cm diameter aluminum sphere into quartz sand in low effective gravities ($\sim0.2 - 1$ m/s$^2$).  Using in-situ accelerometers we measure the acceleration profile during the impacts and determine the peak accelerations, collision durations and maximum penetration depth. We find that the penetration depth scales linearly with the collision velocity but is independent of the effective gravity for the experimental range tested, and that the collision duration is independent of both the effective gravity and the collision velocity. No rebounds are observed in any of the experiments.  Our low-gravity experimental results indicate that the transition from the quasi-static regime to the inertial regime occurs for impact energies two orders of magnitude smaller than in similar impact experiments under terrestrial gravity.  The lower energy regime change may be due to the increased hydrodynamic drag of the surface material in our experiments, but may also support the notion that the quasi-static regime reduces as the effective gravity becomes lower. 
\end{abstract}



\section{Introduction}

\noindent Space missions \citep[\eg][]{sullivan2002,robinson2002,fujiwara2006,veverka2000,coradini2011,jaumann2012} and thermal infrared observations \citep[\eg][]{Campins2009, Gundlach2013, delbo2015} have revealed that asteroids are covered with substantial regolith \citep[the loose unconsolidated material that comprises the upper portions of an asteroid;][]{robinson2002}.  Several current and future small body missions include lander components \eg MASCOT and the MINERVA rovers on-board JAXA's Hayabusa-2 mission \citep{tsuda2013}, MASCOT-2 and possibly AGEX on board ESA's AIM mission \citep{michel2016,Ho2016,karatekin2016}.  Given the small escape velocities of these missions' targets, the landing velocities are likely to be small (10's of cm/s or lower) in order to minimise the risk of rebounding into space.  The understanding of low-velocity surface-lander interactions is, therefore, important for all missions with lander components and will influence the lander deployment strategy, the mission design and operations, and even the choice of payload for the future missions \citep[\eg][]{Murdoch2016EGU}. The dynamics of low-velocity interactions with granular material in reduced gravity are also important for other missions, such as OSIRIS-REx (NASA), that will interact directly with the asteroid's surface in order to retrieve a regolith sample \citep{Lauretta2012}.

In addition to being of high importance for future space missions, the physics of low-velocity collisions in low-gravity also has consequences for our understanding of planetary accretion processes, planetary ring dynamics, cratering processes, asteroid geophysical evolution and our interpretations of small body surfaces. For example, rubble-pile asteroids such as (25143) Itokawa are thought to be formed via a catastrophic disruption event and subsequent re-accumulation \citep[\eg][]{michel01}. The impact velocity during re-accumulation is limited by the escape velocity of the body \citep[$\sim10$ cm/s for Itokawa;][]{fujiwara2006} and, as such, the collisions involved are necessarily at low velocity.  Experimental research can improve our understanding of the different processes that may arise during gravitational re-accumulation such as rebounding or burial of the impacting particle or boulder, regolith mixing, or secondary crater formation \citep{Nakamura2008}. 

Using the ISAE-SUPAERO drop tower \citep{sunday2016}, we have performed a series of low-velocity collisions into granular material in reduced gravity. Reduced gravity is simulated by releasing a free-falling projectile into a surface container with a downward acceleration less than that of Earth's gravity. The acceleration of the surface is controlled through the use of an Atwood machine, or a system of pulleys and counterweights. This system provides a means to reduce the effective surface acceleration of the granular material. Since both the surface and projectile are falling, the projectile requires some time to catch up with the surface before the collision begins. This extended free-fall period increases the experiment duration, making it easier to use accelerometers and high-speed cameras for data collection.  The experiment is built into an existing 5.5 m drop tower frame \citep[originally built for aircraft and material drop-tests; ][]{israr2014} and has required the custom design of all components, including the projectile, surface sample container and release mechanism \citep{sunday2016}.  The design of our experiment accommodates effective accelerations of $\sim$0.1-1.0 m/s$^2$. This is lower than in previous experiments \citep[\eg][]{altshuler14, goldman08}, allowing us to come closer to the conditions found at the surface of asteroids.  

Here we will first discuss previous work in the field of low-velocity granular collisions, before describing the experiment, the data collection and the data analysis. Finally we present the results of our experimental trials and discuss the implications for small body missions and asteroid evolution. 

\section{Low-velocity granular impacts}

\noindent Granular materials exhibit several characteristics that make them interesting but equally very difficult to understand.  Unlike solids, they can conform to the shape of the vessel containing them, thereby exhibiting fluid-like characteristics. On the other hand, they cannot be considered a fluid, as they can be heaped \citep{gudhe94}.  The micro-gravity environment at the surface of an asteroid, in combination with the granular surfaces, challenge existing theoretical models.  In this paper we focus specifically on low-velocity granular impacts, however, for a detailed discussion of granular materials in the context of small body science, including many other applications where understanding granular dynamics in low gravity is important, the reader is referred to \cite{murdochAsteroidsIV2015}. 

As mentioned above, low-velocity impacts into granular material are of interest for many aspects of planetary science (planetary accretion, planetary ring dynamics, cratering, asteroid re-accumulation, regolith mixing, ...) and for current and future space missions that will interact with the surfaces of small bodies. However, the subject of granular impacts is also of great interest to the granular physics community and has been the focus of multiple studies. For example, as the resulting crater depth relates to the stopping force on the impactor, the maximum penetration depth in a granular impact can be used to probe granular mechanics. \cite{uehara2003}, \cite{newhall2003} a,d \cite{ambroso2005a} show experimentally that under terrestrial gravity, the final penetration depth ($z$) of a spherical or cylindrical projectile impacting a dry granular material (glass beads) is related to the drop height ($H$) as $z \sim H^{1/3}$ (or to the collision velocity as $z \sim V_c^{2/3}$). \cite{Tsimring2005} found the same dependance using numerical simulations. The experiments of \cite{goldman08} and \cite{Bruyn2004} on the other hand, demonstrate a linear scaling of penetration depth with impact velocity ($z \sim V_c$) for impacts of spheres into glass beads. This linear scaling with impact velocity was also found in the numerical simulations of \cite{Ciamarra2004}. The different scaling relationships may arise due to different packing fractions of the granular materials \citep{Bruyn2004}, or due to the differences between shallow and deep penetration experiments \citep{uehara2003,walsh2003}. 

The collision time for a spherical impactor is found to be independent of the collision velocity at higher collision velocities \citep{goldman08,Ciamarra2004} but for lower collision velocities ($\gtrsim$1.5 m/s), the collision duration increases with decreasing collision velocity.

The determination of the drag force during penetration was the focus of the work by \cite{Nakamura2013}, who attempt to better understanding the penetration of particles into regolith during the gravitational re-accumulation process of an asteroid following a catastrophic disruption.  To do this, they study the deceleration of spherical plastic projectiles as they impact glass beads at speeds of $\sim$70 m/s. The experiments, performed under both terrestrial gravity and micro-gravity (using a parabolic flight), allowed estimates to be made of the penetration depth of an impactor on an asteroid surface.

With the aim of investigating planetesimal growth and planetary ring dynamics, \cite{Colwell1999} and \cite{colwell03} studied micro-gravity collisions into granular surfaces over the course of two different payload experiments aboard the Space Transportation System (Space Shuttle). In the first set of experiments \citep{Colwell1999} spherical Teflon projectiles of 0.96 cm and 1.92 cm diameter impacted into JSC-1 (a glass-rich basaltic ash similar to lunar mare regolith) at 10 - 100 cm/s in order to study both the coefficient of restitution and the ejecta velocities. They found that virtually no ejecta was produced in these collisions and the coefficients of restitution were very low (0.02 - 0.03). These experiments were later repeated with 2 cm diameter projectiles and less compacted targets of JSC-1 and quartz sand \citep{colwell03}. Again, coefficients of restitution of 0.01 - 0.02 were observed for impacts in the 15 - 110 cm/s range but no rebound was observed for impacts at less than 12 cm/s. This time, however, in the higher velocity impacts ($>$25 cm/s) some ejecta was observed from the loose quartz sand targets.  The extension of this same experiment to even lower impact energies was achieved by \cite{Colwell2015} who completed a series of micro-gravity impact tests over three parabolic flights. These results showed an increase in the coefficient of restitution for marbles impacting JSC-1 at very low impact velocities ($\sim$5 cm/s).  These Space Shuttle and parabolic flight experiments were very successful but the need for future experiments varying the impactor mass and impactor velocity was highlighted \citep{Colwell1999, colwell03}. 

Having accelerometers inside the projectiles allows for a better understanding of the impact dynamics. Such experiments were performed by \cite{goldman08} both under terrestrial gravity and in reduced-gravity making use of an Atwood machine similar to the experiment described here. In their study the acceleration of spherical and disk-shaped projectiles was measured during the impact into granular material (glass beads). They showed that the peak acceleration of a sphere during the impact scales with the square of the impact velocity, however, this scaling does not hold for impact velocities $<$1.5 m/s. Additionally, for spheres, the collision duration is found to be independent of the collision velocity at impact velocities $>$1.5 m/s, but at lower impact velocities the collision duration increases with decreasing impact velocity. The opposite effect is seen for an impacting disk, with the collision duration decreasing significantly at low impact velocities. 

\cite{altshuler14} also use an Atwood machine to study impacts of spheres into polystyrene beads in varying gravity. They found that the maximum penetration depth of the projectile into the granular material was independent of the gravitational acceleration. Also, in addition to the finding of \cite{goldman08} that the collision time increases with decreasing velocity for a spherical impactor, \cite{altshuler14} find that the collision duration also increases with decreasing gravity. It should be noted, however, that the extrapolation to very low gravity levels ($<$ 0.4 m/s $^2$) was made using numerical simulation data and not experimental data. 

These last two studies \citep{altshuler14,goldman08} have started to investigate the role of gravity during low-velocity granular impacts. However, the experiments have collision velocities of $\sim$40 to 700 cm/s and gravity levels $>$0.4 m/s$^2$.  Here we use the ISAE-SUPAERO drop tower experiment \citep{sunday2016} in order to achieve lower and variable gravity levels, lower collision velocities, and to increase the size of the experiment. The scale of our experiment allows us to use a larger projectile (10 cm; closer in size to an asteroid lander) that contains accelerometers in order to study in detail the impact dynamics during the low-velocity, low-gravity collisions.  As a secondary data source, we also use a static rapid, high resolution camera. Also, rather than using glass or polystyrene beads, we use quartz sand in our experiment in order to be more representative of the regolith found on small bodies.

\section{Experiment Details}

\noindent The detailed experiment design is described in \cite{sunday2016}. Here, we just give a short overview of the key aspects. The acceleration of the surface is controlled through the use of an Atwood machine, or a system of pulleys and counterweights, which allows the surface container to have a constant downward acceleration less than that of gravity.  If pulley friction and chord elasticity are neglected, then the controlled acceleration is simply a function of mass. The expression for the surface container's acceleration ($a_s$) is derived by balancing the forces on the surface container and counterweights. This is given by Equation \ref{eq:cwMass}, where $m_s$ is the mass of the surface container, $m_{cw}$ is the total combined mass of all counterweights, and $g$ is the Earth's gravitational acceleration. 

\begin{equation} 
	\centering
	{a_{s} = g\left(\frac{m_{s} - m_{cw}}{m_{s} + m_{cw}}\right)}
	\label{eq:cwMass}
\end{equation}

\noindent  The effective gravity of the surface container ($g_{eff}$) is then the difference between the Earth's gravitational acceleration and the acceleration of the surface container \ie $g_{eff} = g - a_s$. 

\begin{figure}
	\centering
	\includegraphics[scale=1]{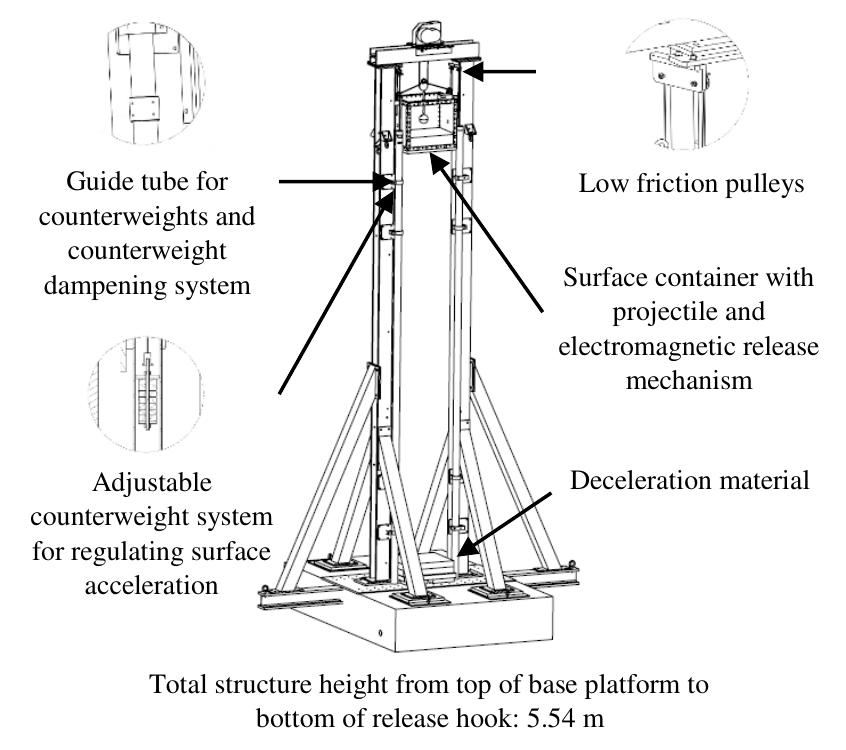}
    \includegraphics[scale=1]{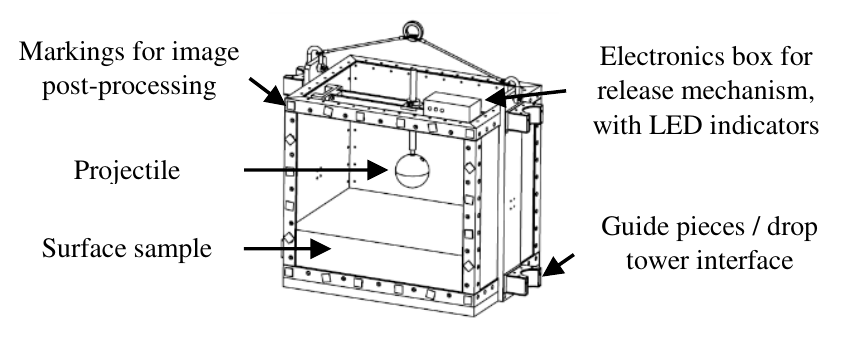}
	\caption{{\bf Isometric line drawing of the experiment and surface container.} Isometric line drawing of the (upper figure) drop tower structure and (lower figure) surface container sub-assembly. The experiment's custom-designed subsystems include the surface container (outer dimensions: 62 cm long, 45 cm wide, and 59 cm high), projectile (10 cm diameter), electromagnetic release mechanism, pulley and counterweight system, and deceleration material.  The reference markings for image post-processing can also be seen on the surface container. Images from \protect\citep{sunday2016}. }
	\label{fig:fullAss}
\end{figure}

The experiment is built into an existing 5.5 m drop tower frame and has required the custom design of all components, including the projectile, surface sample container and release mechanism. The counterweight holders alone weigh 400 grammes each. Then, mass can be added to the holders at 100 gramme increments in order to change the acceleration of the surface container. \Fig{fullAss} shows the mounted pulley system, and the counterweight, and guide tube components of the assembly.

The surface container sub-assembly comprises of three parts: the surface container, the release mechanism, and the projectile.  As described in \cite{sunday2016}, the surface container is sized so that, for a 10 cm projectile, the walls of the container will not influence the rebound dynamics of the collision and is 62 cm $\times$ 45 cm $\times$ 59 cm in size.  The front and back panels of the surface container are made of 10 mm thick Makrolon polycarbonate material, while the two side panels are made of a light-weight aluminum alloy (4 mm thick). A narrow beam traverses the center of the container and acts as a support for the electromagnetic release mechanism. An electromagnet is mounted at the end of a supported tube, which can be raised and lowered to change the separation distance between the projectile and the surface.  The electronics box for controlling the electromagnet is mounted to the top of the container.  \Fig{fullAss} shows an illustration the surface container and the location of its different features. The total mass of the container assembly, including approximately 80 kg of sand, is 160 kg. 

\begin{figure*}
	\centering
	\includegraphics[width=0.4\textwidth]{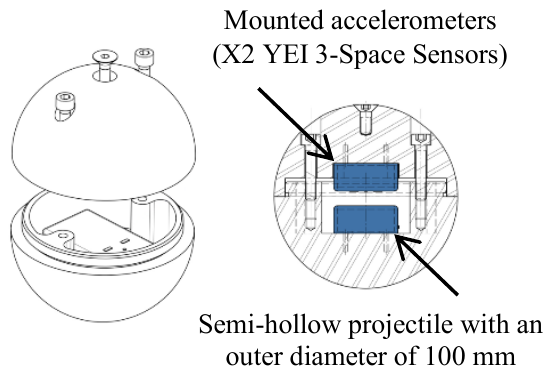} \\
		\includegraphics[width=0.2\textwidth]{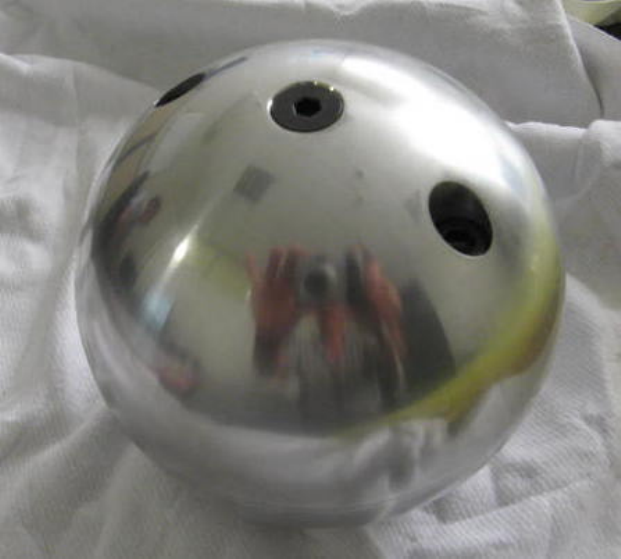}
	\includegraphics[width=0.335\textwidth]{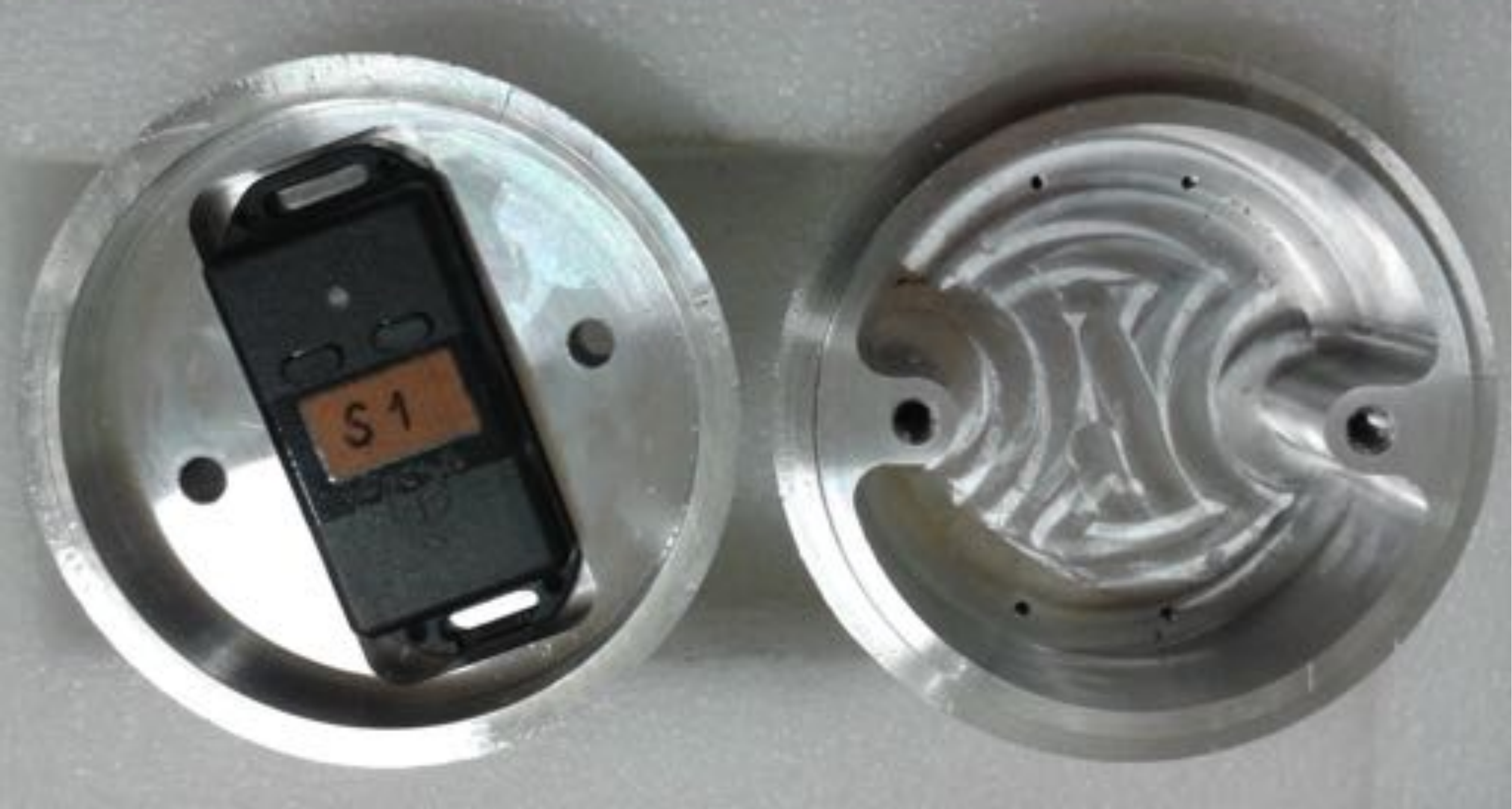}
	\caption{{\bf Projectile.} (Above) Line drawing of the semi-hollow experiment projectile, with two mounted YEI 3-Space Sensors \protect\citep{sunday2016}. (Lower left) Photograph of the closed projectile showing the metallic screw on the top that attaches to the electromagnet. (Lower right) Photograph of the open projectile with one YEI 3-Space Sensor attached.}
	\label{fig:projectile}
\end{figure*}

The 10 cm diameter spherical projectile used in these experiments, shown in \fig{projectile}, is specifically designed to accomodate two wireless accelerometers (see \sect{sensors}) and to have the centre of gravity at the centre of the sphere. It is fabricated out of 2017 aluminum alloy and weighs 1000 grammes (1056 grammes including the two accelerometers).

\subsection{Sensors}
\label{s:sensors}

\subsubsection{Accelerometers}

YEI 3-Space Sensors \citep{YEI} are mounted to the projectile and surface container. These sensors are data-logging devices that contain an Attitude and Heading Reference System (AHRS), an Inertial Measurement Unit (IMU), and a micro-SD card for on-board data storage. The YEI 3-Space Sensors were specifically selected because of their low mass, high sensitivity and high shock resistance (see \tbl{YEI}).  These features allow the sensors to record the impact between the projectile and the sand with high precision and to survive to the final shock at the end of the drop. Several data filtering options are available as part of the sensors. We use only the data logging capability with no automatic filtering giving the maximum sampling frequency possible of $\sim$1200 Hz. 

\begin{table}
\centering
\caption{YEI 3-Space Sensor characteristics. Data from \protect\cite{YEI}.}
\label{t:YEI}
\begin{tabular}{ll}
Dimensions & 35 mm $\times$ 60 mm $\times$ 15 mm \\ 
Weight & 28 grammes \\ 
Shock survivability & 5000$g$ \\ 
Accelerometer scale & $\pm2g$ / $\pm4g$ / $\pm8g$ selectable \\ 
Accelerometer resolution & 14 bit \\ 
Accelerometer noise density & 99$\mu g /\sqrt Hz$  \\ 
& 0.00024$g$/digit for $\pm2g$ range \\
Accelerometer sensitivity  & 0.00048$g$/digit for $\pm4g$ range \\
& 0.00096$g$/digit for $\pm8g$ range \\ 
\end{tabular}
\end{table}

\subsubsection{Cameras}

A high resolution camera (Ultima APX-RS Photron FASTCAM) is used with a Sigma 24-70mm f/2.8 DG lens, to capture high-speed images (1,000 frames per second) of the projectile-sand collision with a 1024 x 1024 pixel resolution. The camera was static and was placed at a distance of $\sim$2.7 m from front of the surface container and a focal length of 24 mm was used for the lens. This gives a pixel resolution of $\sim$5 pixels per cm at the front panel of the surface container.  In addition, a small, wide angle (175$^\circ$) camera (PNJ AEE MagiCam SD100) was fixed inside the surface container to give an in-situ view of the experiment. This camera has a pixel resolution of 1920 $\times$ 1080 and captures images at 30 frames per second. 

\subsection{Surface material}
\noindent 
The surface material used in these trials was quartz sand (98.7\% SiO$_2$) with a size range of 1-2.5 mm and a median grain diameter (the grain diameter for which half the sample by weight is smaller and half is larger) of 1.83 mm. The detailed granulometric information is provided in \tbl{granulometry}.  With this size range of particles, we do not expect to be sensitive to interstitial air effects; these have been shown to be negligible for grains with diameters $>$0.1 mm \citep{Katsuragi2016, Pak1995}. This is demonstrated specifically for low-speed impacts into a granular material by \citep{katsuragi07}. The individual grain density is 2.65 g/cm$^3$ and each grain has a measured hardness of 7 Mohs \citep{FibreVerte}. 

In order to measure the angle of repose, sand was allowed to pour from a bottle onto a flat surface. The surface was covered with the same sand (glued to the surface). The experiment was filmed and repeated three times.  Using the final image, and the measuring tool in GIMP, the angle of repose of the sand on both sides of the pile was measured for each trial. The resulting angle of repose was found to be 32.5$^\circ$ $\pm$ 2.4$^\circ$. This is a very a typical value for sand in a terrestrial environment. Lunar soil samples have been found, however, to have angles of repose ranging from 25$^\circ$ - 50$^\circ$ \citep{carrier1991}; the higher values are likely due to the more angular particles that arise from fragments generated from impact comminution. Lunar regolith simulants, such as JSC-1, therefore contain more angular particle shapes and have a higher angle of repose than our quartz sand ($\sim45^\circ$; \citep{McKay1994}).  Martian regolith simulants, such as SSC-1 and SSC-2, have also been developed and these have angles of repose ranging from 35$^\circ$ to 41$^\circ$ \citep{scott2009}. On asteroids, gravitational slopes above angles of $\sim35^\circ$ are rare \citep{scheeres06b,thomas02}. This indicates that the angle of repose may be lower for asteroid regolith, despite the Hayabusa sample return analyses showing that Itokawa regolith particles are more angular than lunar regolith \citep{tsuchiyama2011}. 

The surface container was filled with $\sim$80 kg of sand, reaching a height of $\sim$17 cm. This gives an approximate bulk density of 1790 kg/m$^3$. As noted in \cite{sunday2016}, the deceleration system naturally regulates the bulk density of the surface material, and the bulk density does not change between trials. The sand was brushed, however, to restore a level surface before each trial. 


\begin{table}
\centering
\caption{Sand granulometry provided by FibreVerte. The values were determined from sieve tests, conform to the French national organisation for standardisation  (Association Francaise de Normalisation; AFNOR). Data from \protect\cite{FibreVerte}.}
\label{t:granulometry}
\begin{tabular}{cc}
Mesh opening ($\mu$m) & Cumulative mass (\%) \\ \hline
	$>$ 3150 & 0 \\ 
	$>$ 2500 & 2.2 \\ 
	$>$ 2000 & 31.9 \\ 
	$>$ 1600 & 75.2 \\ 
	$>$ 1250 & 95.3 \\ 
	$>$ 1000 & 98.4 \\ 
	remaining & 1.6 \\
\end{tabular}
\end{table}

%

\newpage
\section{Experimental trials}

\subsection{Typical experiment data}
\begin{figure*}
	\centering
	\includegraphics[width=0.47\textwidth]{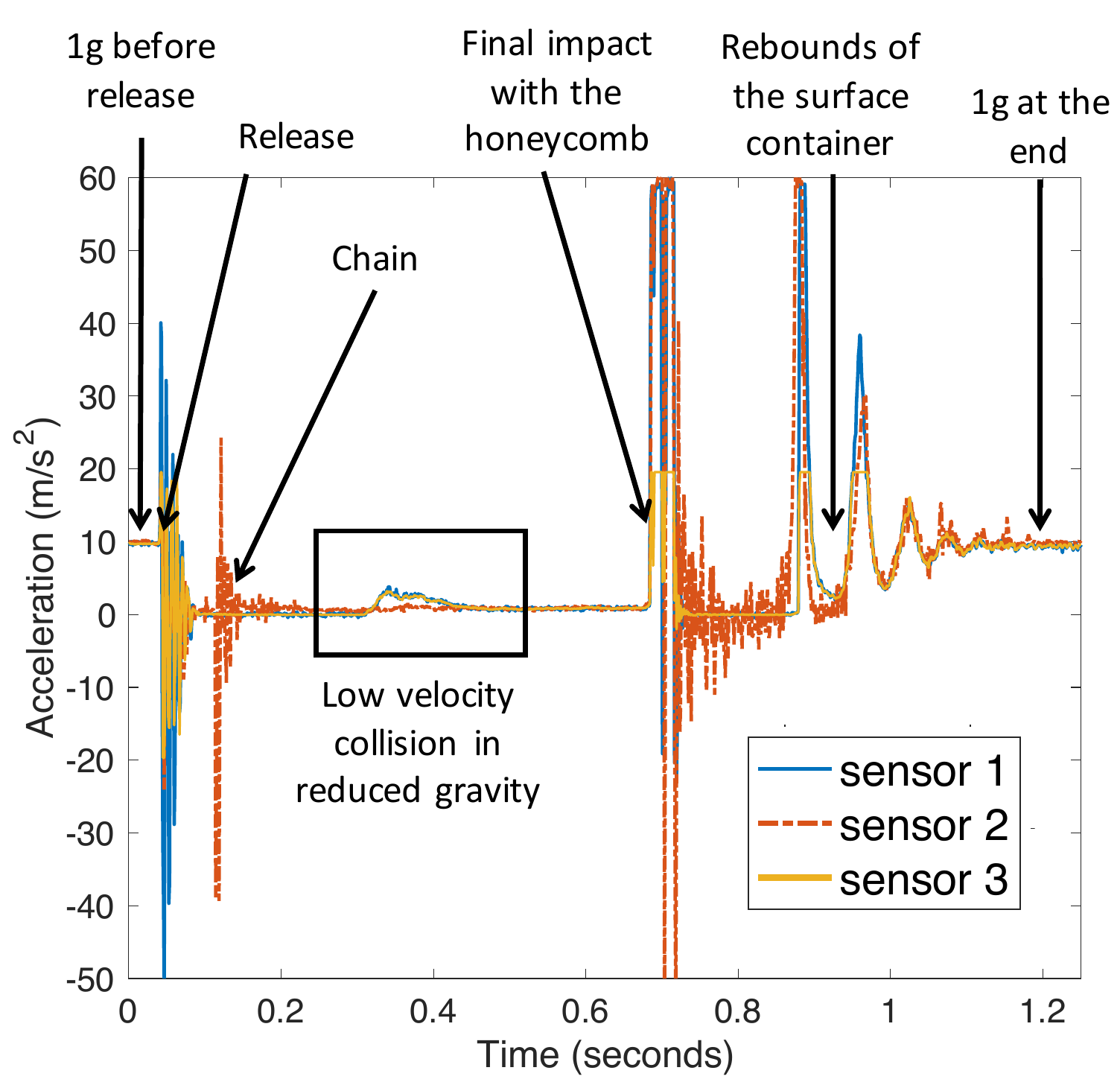} 
	\includegraphics[width=0.52\textwidth]{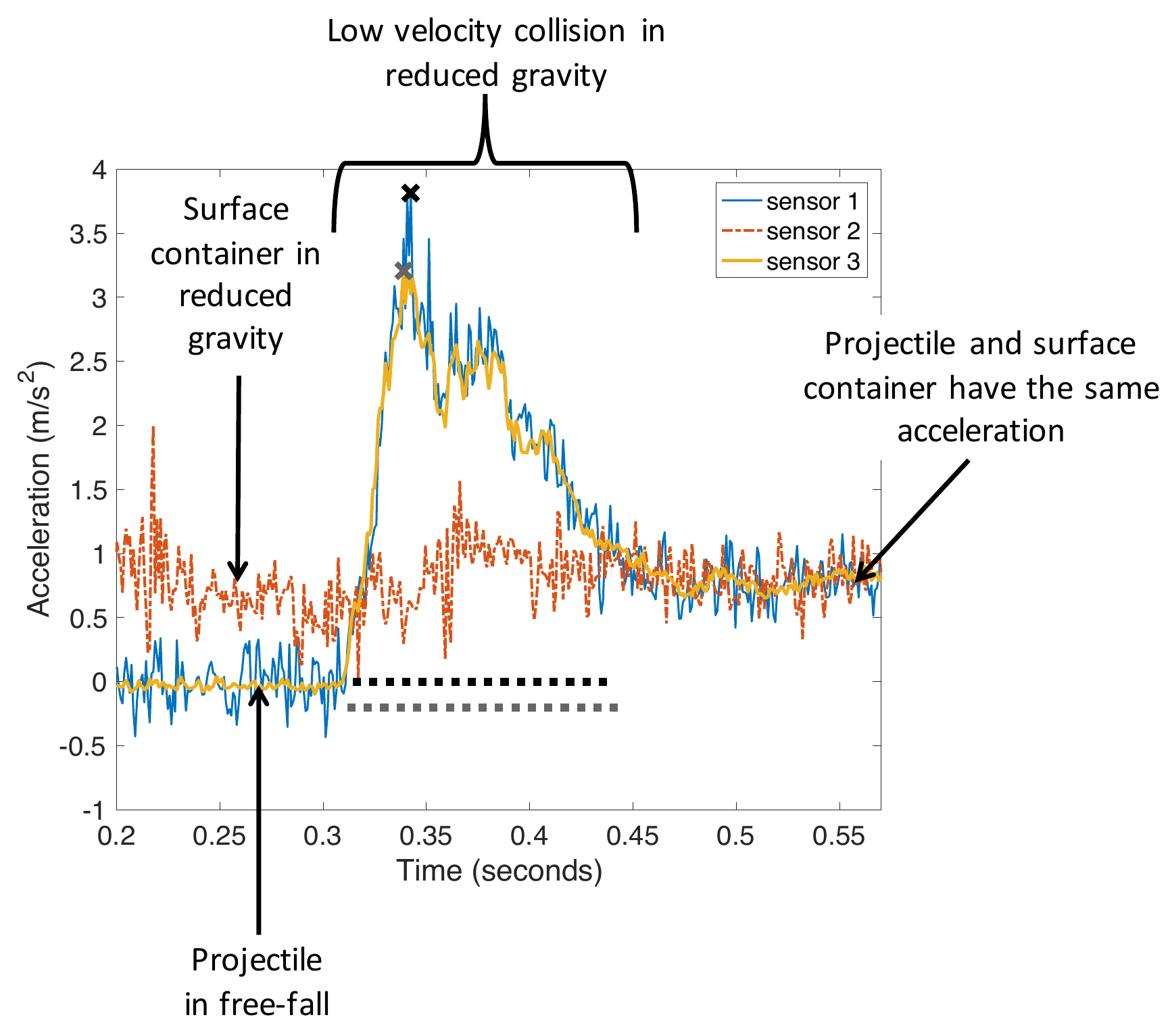} 
	\caption{{\bf Typical raw experiment data.} Left: The acceleration recorded by the three sensors during the entire experiment. Right: Close-up of the accelerometer data during the projectile - sand collision. In both figures the vertical acceleration measured by the three sensors is shown as a function of time. In this example sensors 1 (thin blue line) and 3 (thick orange line) were inside the projectile and sensor 3 (dashed red line) was attached to the surface container.  Sensors 1 and 2 had a dynamic range (scale) of $\pm6g$, and sensor 2 had a dynamic range of $\pm2g$. No filtering or smoothing has been applied to the data. The mean acceleration of the surface container during the collision (the effective gravity) is 0.81 m/s$^2$. The peak accelerations measured by sensors 1 and 3 during the collision are shown by the black and grey crosses, respectively. Similarly, the collision duration calculated from each of the projectile sensors is indicated by the black (sensor 1) and grey (sensor 3) horizontal dashed lines}
	\label{fig:typical}
\end{figure*}

\begin{figure*}
	\centering
	\includegraphics[width=1\textwidth]{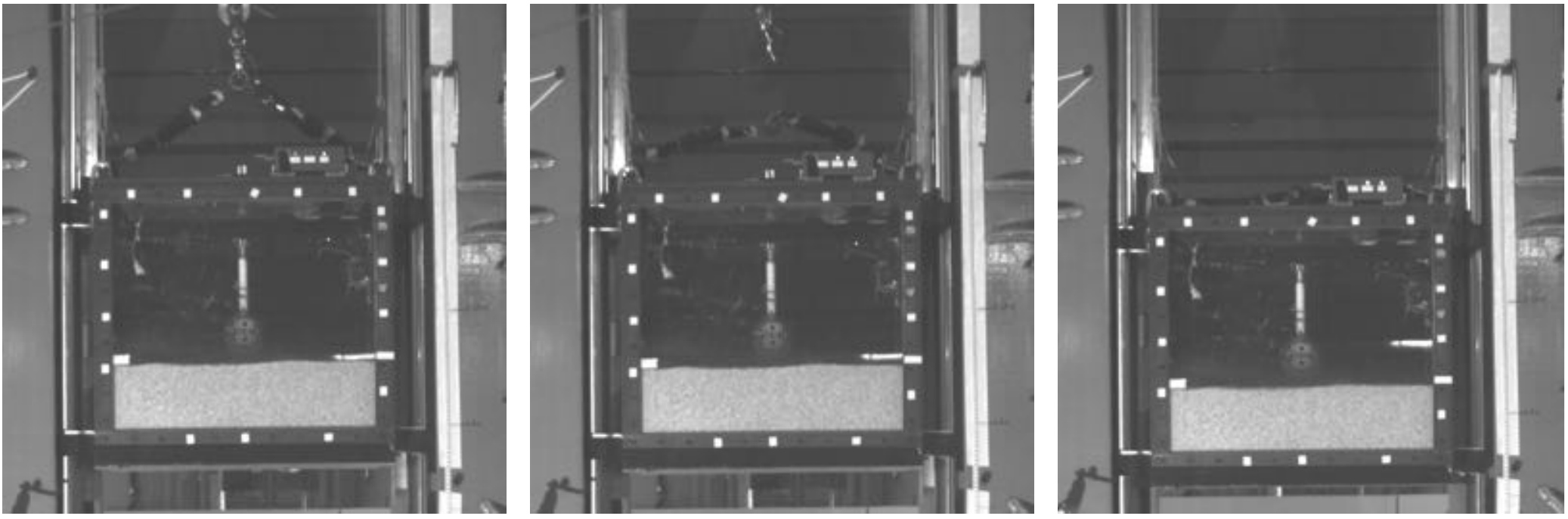} 
	\caption{{\bf Chain falling onto surface container.} Before release the surface container is held by a large chain (left image). Upon release, the surface container begins to fall but is slowed down by the counterweights (middle figure). The chain, accelerating faster than the surface container, falls onto the surface container (right figure) causing the vibrations seen at $\sim$0.1 s in \fig{typical}. The time delay between release and the chain hitting depends on the acceleration of the surface container.}
	\label{fig:chain}
\end{figure*}

\noindent 
Shown in \fig{typical} is a set of typical experiment data. In this example, the initial separation distance between the sand surface and the projectile was 2 cm and there were 4.8 kg of counterweights (including the mass of the counterweight holders). Sensors 1 and 3 were in the projectile, and sensor 2 was attached to the container.  No filtering has been applied to the data.  At the start of the experiment, before the container release, the accelerometers all measure 1$g$. At the moment of release there are extensive vibrations due to the mechanical release mechanism \citep[for details, see][]{sunday2016}. These vibrations are larger for the projectile as, being attached only by the electromagnetic, it is free to move. After approximately 0.1 s there is a large signal recorded by the sensor attached to the surface container. This is due to the chain falling onto the surface container (\fig{chain}).  Next, close to 0.4 s, the low velocity collision between the projectile and the sand in reduced gravity can be seen. Finally, at 0.7 s the surface container impacts the honeycomb material and the container and projectile rebound until coming to rest.

A good experiment, such as the \fig{typical}, is one in which the following criteria are met: (1) the chain falls onto the surface container a sufficiently long time before the projectile - sand collision, (2) there are no large vibrations of the surface container due to friction during the  projectile - sand collision and, (3) the projectile comes to rest on the surface of the sand before the surface container impacts the honeycomb material. The last point, demonstrated by the projectile and surface container having the same acceleration (see \fig{typical}, right) indicates that the collision has entirely finished within the period of time of the drop. 

In the experiment data we typically see fluctuations in the projectile acceleration during impact. Some of these fluctuations can be attributed to the accelerometer noise, or the variable acceleration of the sand container. However, others, are likely due to the creation and annihilation of elements of the force network in the granular material as reported in \cite{goldman08}.  

\section{Data Analysis}

The primary data source for these experiments is the accelerometers. However, we also make use of the high-resolution images in order to validate the accelerometer analyses. The different processes involved are described below.

\subsection{Synchronisation of the accelerometers}
\begin{figure}
	\centering
	\includegraphics[width=0.45\textwidth]{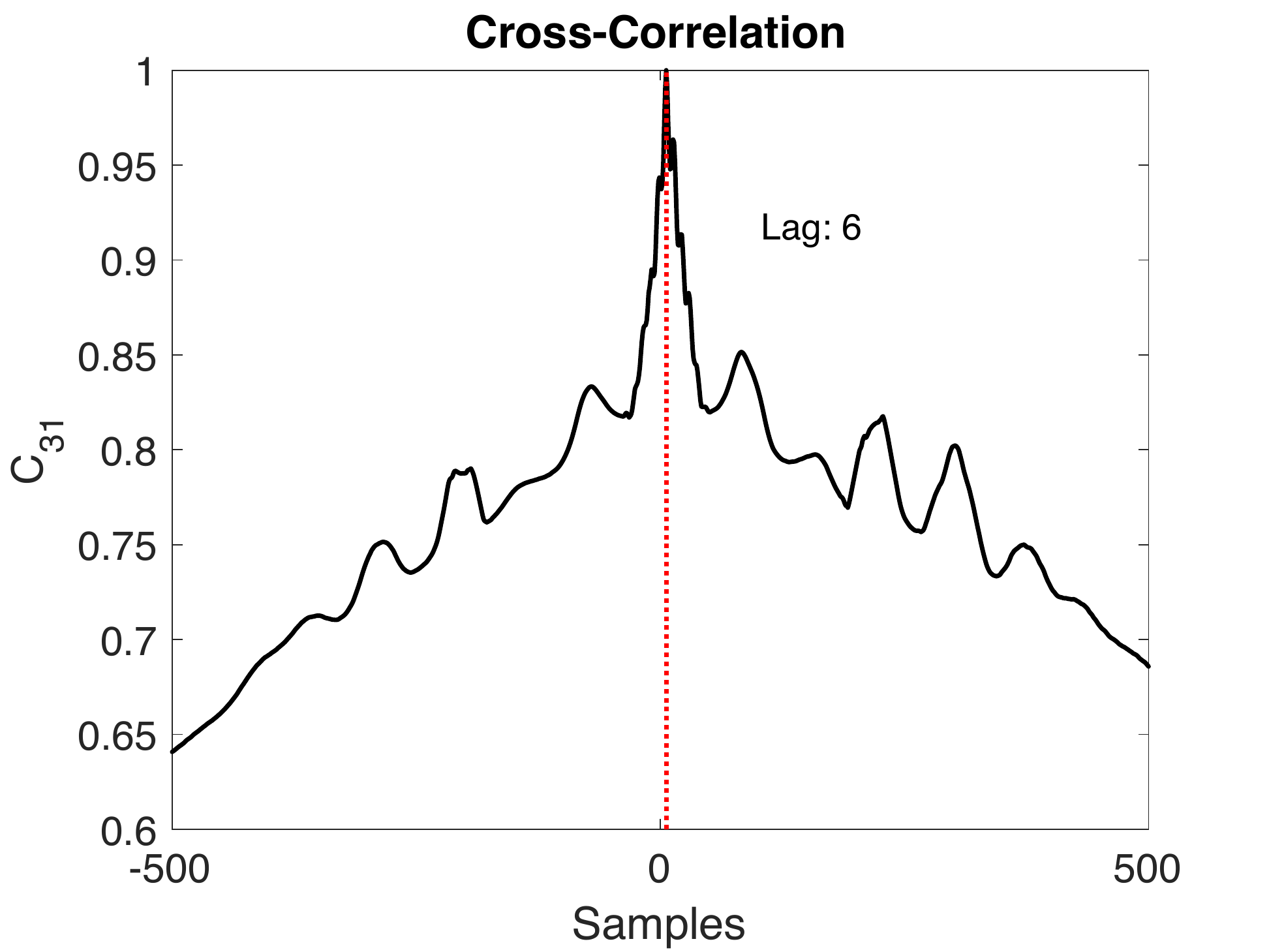} 
	\caption{{\bf Cross correlation of the two projectile sensors. }The location of the maximum value of the cross-correlation indicates the time lead or time lag. In this example there is a time difference of 6 samples (5 ms) before the fine synchronisation is applied.}
	\label{fig:cross}
\end{figure}

\begin{figure}
	\centering
	\includegraphics[width=0.4\textwidth]{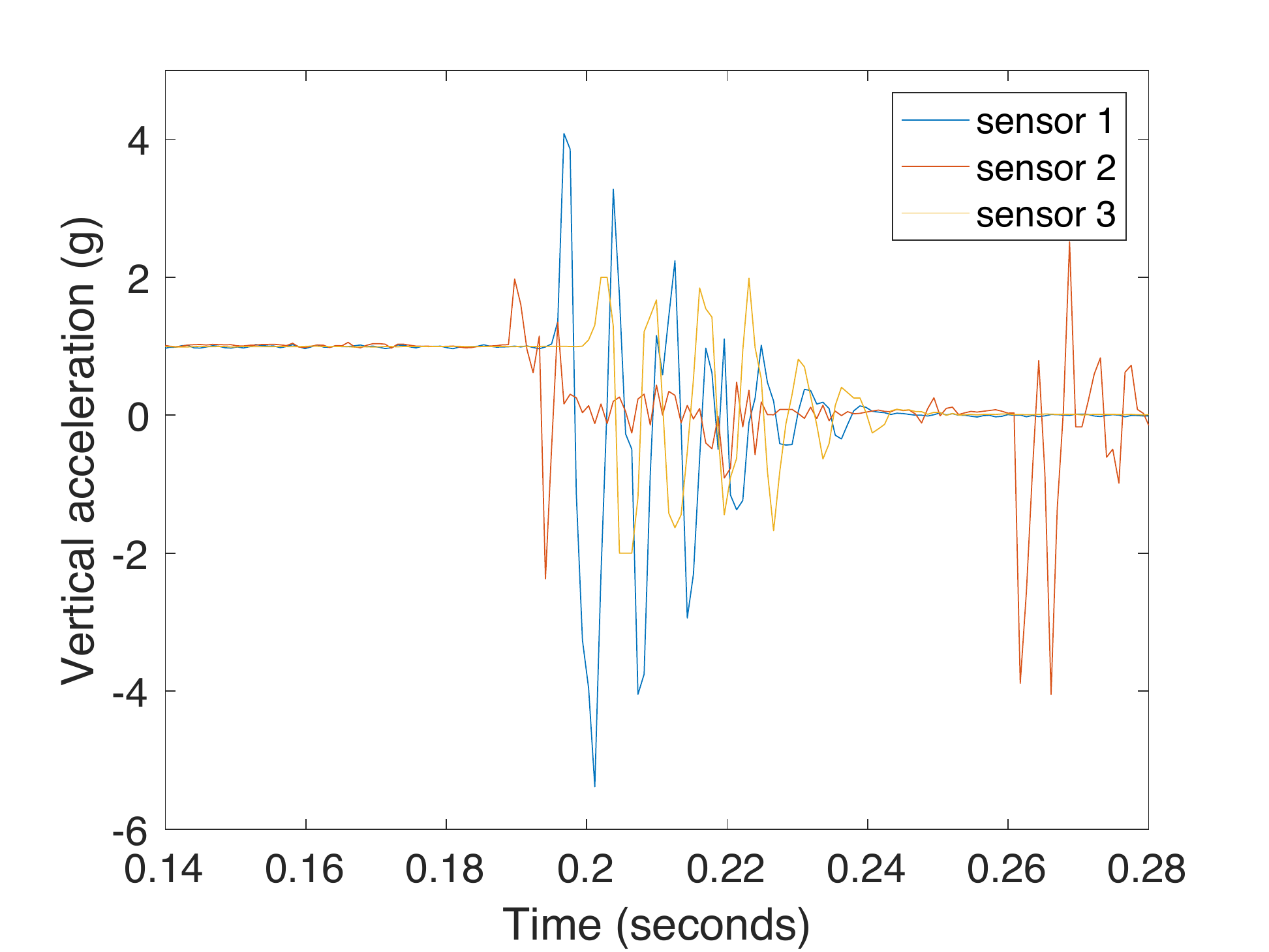} 	
	\includegraphics[width=0.4\textwidth]{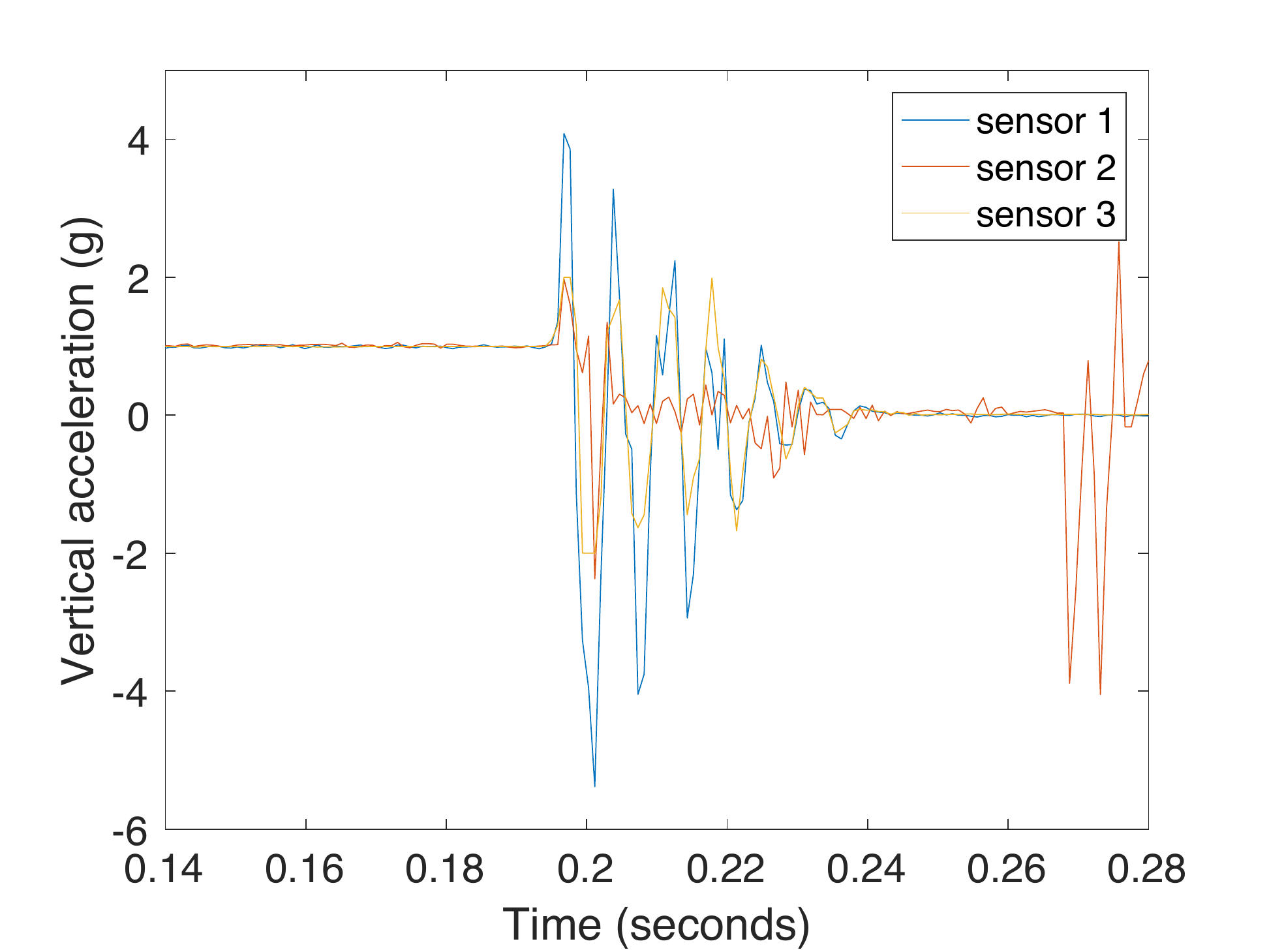} 
	\caption{{\bf Synchronisation of the sensors. } Left: Vertical acceleration recorded by the three sensors following a rough synchronisation performed by hand with a zoom on the moment of release showing that the data are not perfectly synchronised. Right:  Vertical acceleration recorded by the three sensors following the fine synchronisation performed via cross-correlation, also with a zoom on the moment of release, showing that the data are now synchronised. In this example sensors 1 (blue) and 3 (orange) were in the projectile and sensor 3 (red) was attached to the container. }
	\label{fig:synchro}
\end{figure}

\noindent 
First, the drop was identified by hand in the accelerometer data and a $\sim$2 second time period around the drop was extracted from the data of each sensor.  As the sensors were not all started exactly simultaneously, a synchronisation is necessary. The initial, rough synchronisation of the accelerometers was performed by hand using the moment of release of the sand container (top figure, \fig{synchro}). Then, a precise synchronisation was performed automatically. This involved computing the normalised cross correlation of the data between pairs of sensors (\fig{cross}). The location of the maximum value of the cross-correlation indicates the time lead or time lag that is used to align the time vectors of the two sensors. The synchronisation is performed first between the two sensors in the projectile (in the cases where there are two), then the synchronisation is performed between one of the projectile sensors and the surface container sensor. The resulting synchronisation is precise to within 2 samples \ie $<$ 2 ms (lower figure, \fig{synchro}).

\subsection{Collision investigation}
\noindent 
The approximate collision location is identified in the synchronised accelerometer data.  The mean effective acceleration of the sand container during the projectile-sand collision ($g_{eff}$) is then given by the mean of the sand container acceleration over this period.  The peak acceleration ($A_{max}$) of the projectile is the maximum of the projectile sensor acceleration during the collision with the sand. In \fig{typical} (right) the peak acceleration of the two projectile sensors is indicated by the black and grey crosses. 

The moving average of the sand container and projectile acceleration are then calculated as a function of time during the identified time period ($\overline{A}_{box}$ and $\overline{A}_{proj}$, respectively). The start of the collision is defined when $\overline{A}_{proj} > 1.5 \overline{\sigma_{\overline{A}_{box}}}$, where $\overline{\sigma_{\overline{A}_{box}}}$ is the mean standard deviation of the projectile sensor during free-fall. The end of the collision is defined when $\overline{A}_{proj} < \overline{A}_{box} + Y \sigma_{\overline{A}_{box}}$, where $\sigma_{\overline{A}_{box}}$ is the standard deviation of the moving average of the sand container acceleration and $Y$ varies between 0.35 and 1.5 depending on the ratio of the peak acceleration to the mean effective acceleration of the sand container. In \fig{typical} (right) the collision duration calculated from each of the projectile sensors is indicated by the black and grey horizontal dashed lines. The collision duration ($\tau_c$) is then the difference between the collision start and end times.

The vibrations measured during the sand container release can be large enough to saturate the projectile accelerometer(s) as can be seen clearly for sensor 3 in \fig{typical} (left). Due to this, and the fact that the falling chain provides additional vibrations on the sand container accelerometer, it is not possible to simply integrate the accelerometer data of each sensor from the start of the experiment in order to find the relative velocity between the projectile and the sand at the time of the collision. Instead a different approach is used.

\begin{figure*}
	\centering
	\includegraphics[width=1.0\textwidth]{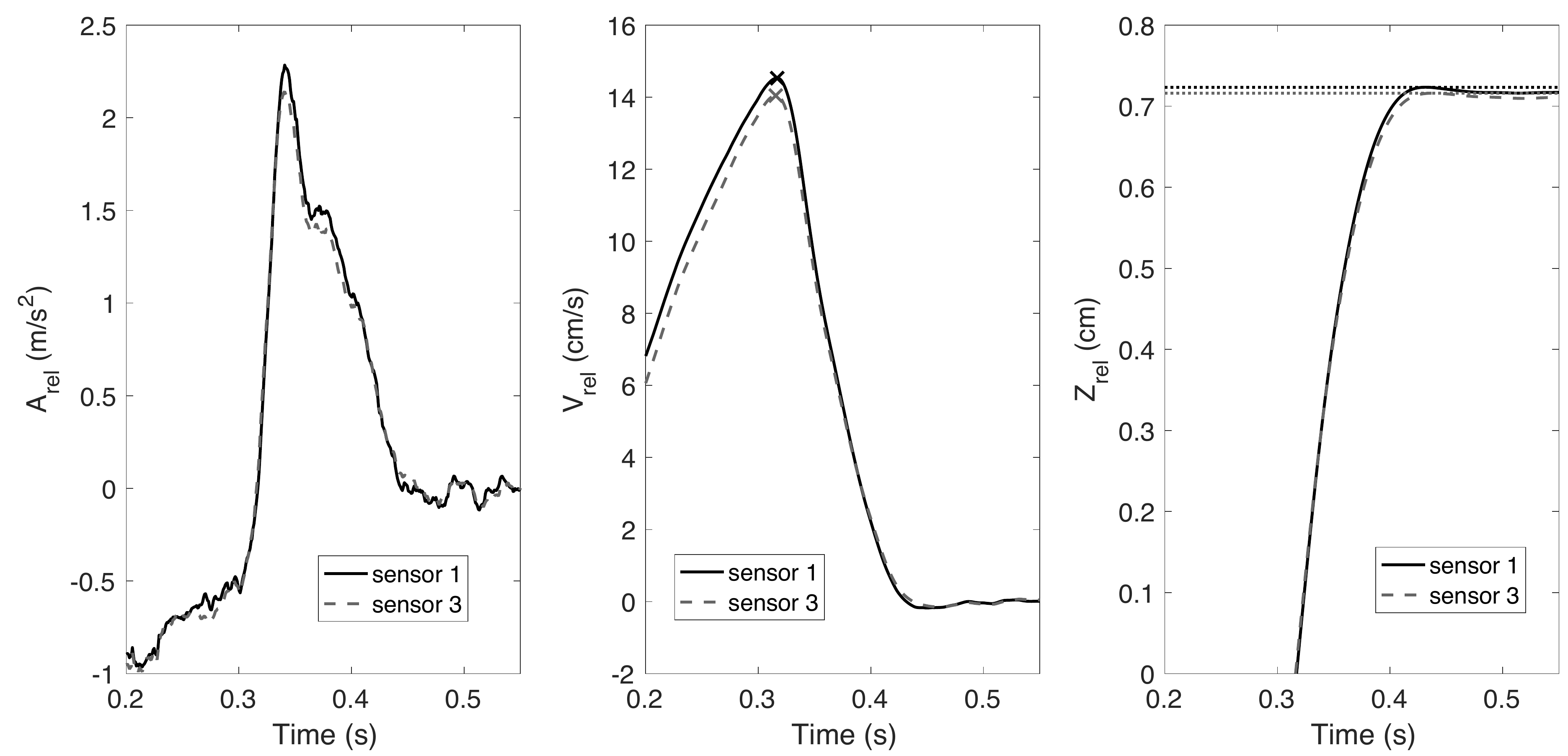} 
	\caption{{\bf Calculation of the collision velocity and penetration depth.} Left: The relative acceleration of the projectile and the sand container for the period around the projectile - sand collision. Middle: The relative velocity of the projectile and the sand container for the period around the projectile - sand collision. The maximum values of the relative velocity - the collision velocity - are indicated by the two crosses. Right: The relative displacement of the projectile and the sand container during the collision \ie the penetration depth of the projectile into the sand. The maximum penetration distances, as measured by the two sensors, are shown by the dotted lines. The two lines in each figure correspond to the relative values between the sand container sensor (\#2) and the two projectile sensors (\#1 and \#3), as shown in \fig{typical}.}
	\label{fig:relativevals}
\end{figure*}

The relative acceleration of the projectile and the sand container is calculated from the moving average accelerations of the projectile and sand container, for the period around the projectile - sand collision ($A_{rel} = \overline{A}_{proj} - \overline{A}_{box}$; \fig{relativevals} left). The relative acceleration is integrated to give the relative velocity of the projectile and the sand container ($V_{rel}$). As the projectile remains in contact with the sand at the end of the collision, the relative velocity at the end of the collision should be zero. This information allows the initial relative velocity at the beginning of the time period to be established and thus the relative velocity throughout the collision (\fig{relativevals} middle). As the projectile immediately starts to slow down upon impact, the collision velocity ($V_{c}$) is the maximum value of the relative velocity (indicated by crosses in \fig{relativevals} middle).  The relative velocity is then integrated, starting from the instant that the collision starts (\fig{relativevals} right). This gives the relative displacement of the projectile and the sand container during the collision and is analogous to the penetration depth of the projectile into the sand ($Z_{rel}$). The maximum relative displacement is, therefore, the maximum penetration depth of the projectile ($Z_{max}$, indicated by the dotted lines in \fig{relativevals} right). The final penetration depth is generally slightly smaller than the maximum penetration depth, due to a slight relaxation of the impact crater following the impact (this can be seen in\fig{relativevals} right).

In the trials where there are two sensors in the projectile, such as \fig{relativevals}, the reported values of peak acceleration, collision duration, collision velocity and maximum penetration depth are the mean values of the two measurements and the reported uncertainty is the standard deviation of the two measurements. 

\subsection{Validation with image analysis}
\noindent 
We make use of the images acquired by the high-speed, high resolution camera to validate the absolute accelerometer measurement, taking the sand container motion as a test case. The image processing flow is based on matrix detection (feature-based tracking). Rather than tracking only one object in the images, several reference makers are tracked meaning than the results are less sensitive to small inaccuracies in the matrix identification and to pixellisation effects. 

%

\subsubsection{Marker identification and tracking }

\begin{figure*}
	\centering
	\includegraphics[width=0.6\textwidth]{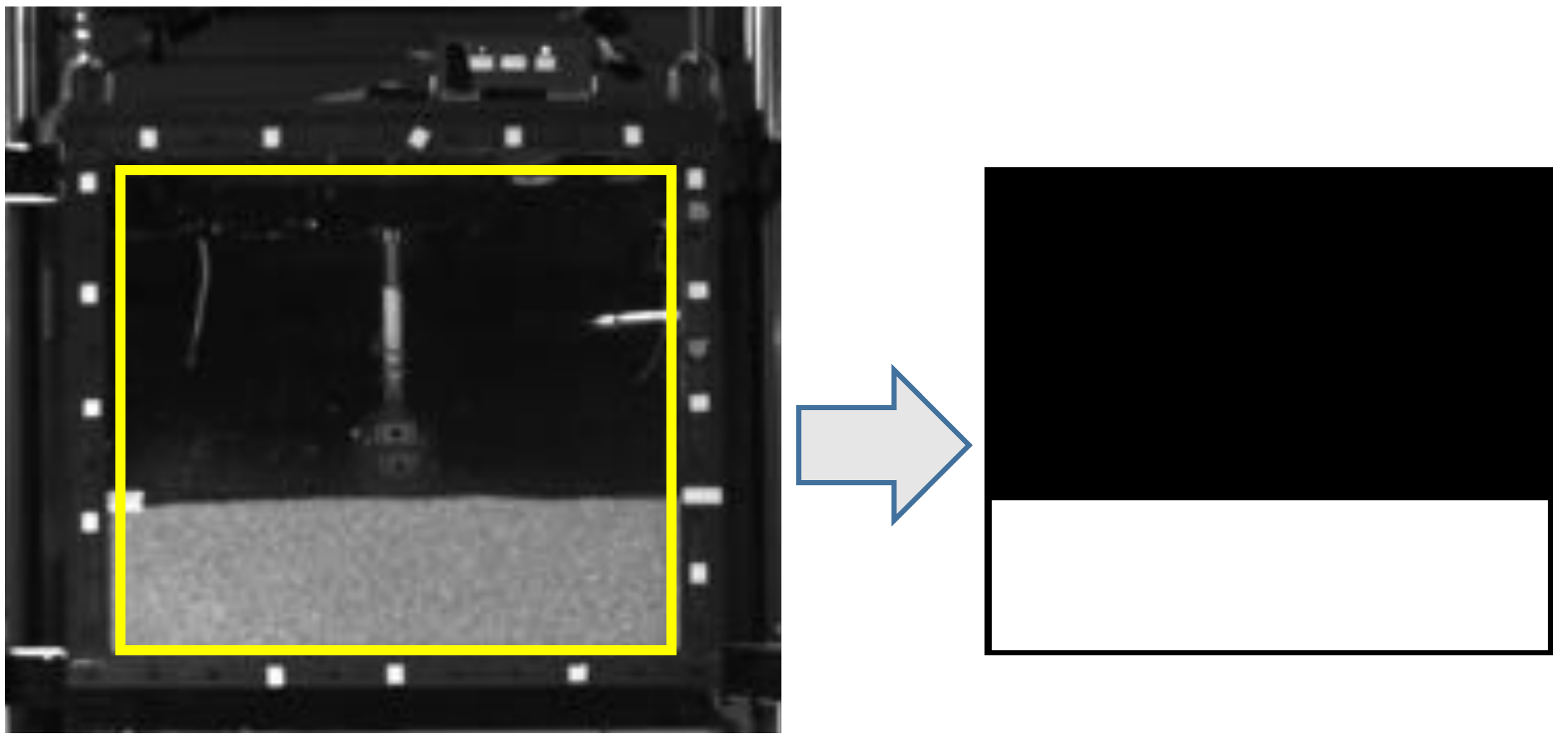}
	\caption{{\bf Sand container model matrix.} The model matrix representing the sand and container visible through the Makrolon front panel. The model matrix has values of 255 (representing the white colour) for the part filled with sand, and 0 representing the black background.}
	\label{fig:sandboxmodel}
\end{figure*}

As a first step, a model matrix representing the sand and container visible through the Makrolon front panel is defined. The model matrix has values of 255 (representing the white colour) for the part filled with sand, and 0 representing the black background, as seen in \fig{sandboxmodel}. The central column of each image is then divided into sub-matrices with the same dimensions as the model matrix and the Minimum Least Square Error (MLSE) is calculated between the model matrix and each sub-matrix. The sand container can then be identified in the image as the sub-matrix with the minimum MLSE. 

The sand container is automatically identified using this method in every 50th image (the full frame rate is not necessary for this first step).  The position data of the sand container are then used to generate a dynamical model for the sand container's motion. As the sand container acceleration is not constant, the acceleration has been modelled as a second order equation (as the friction varies with the square of the speed) and the speed and the displacement are a third and fourth order equations.

\begin{figure*}
	\centering
	\includegraphics[height=0.3\textwidth]{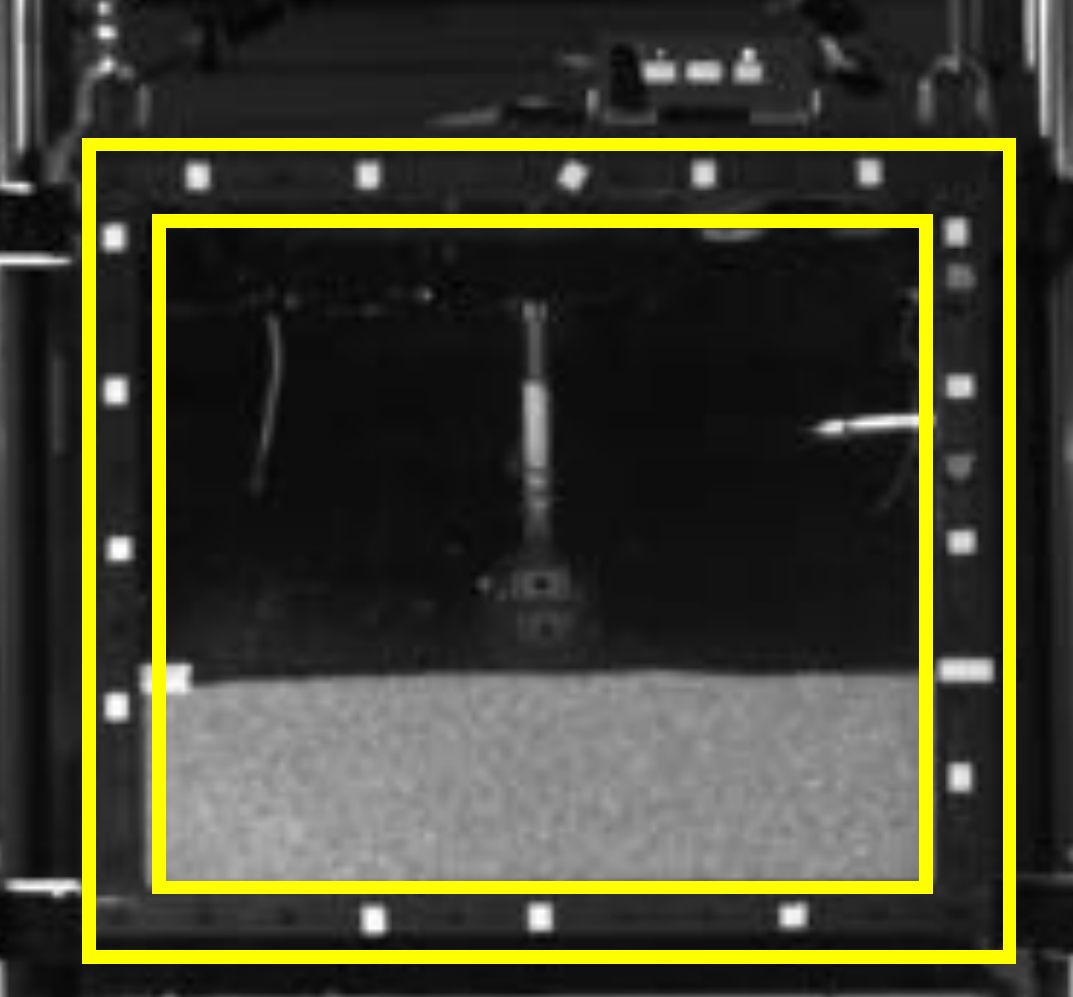}
	\includegraphics[height=0.3\textwidth]{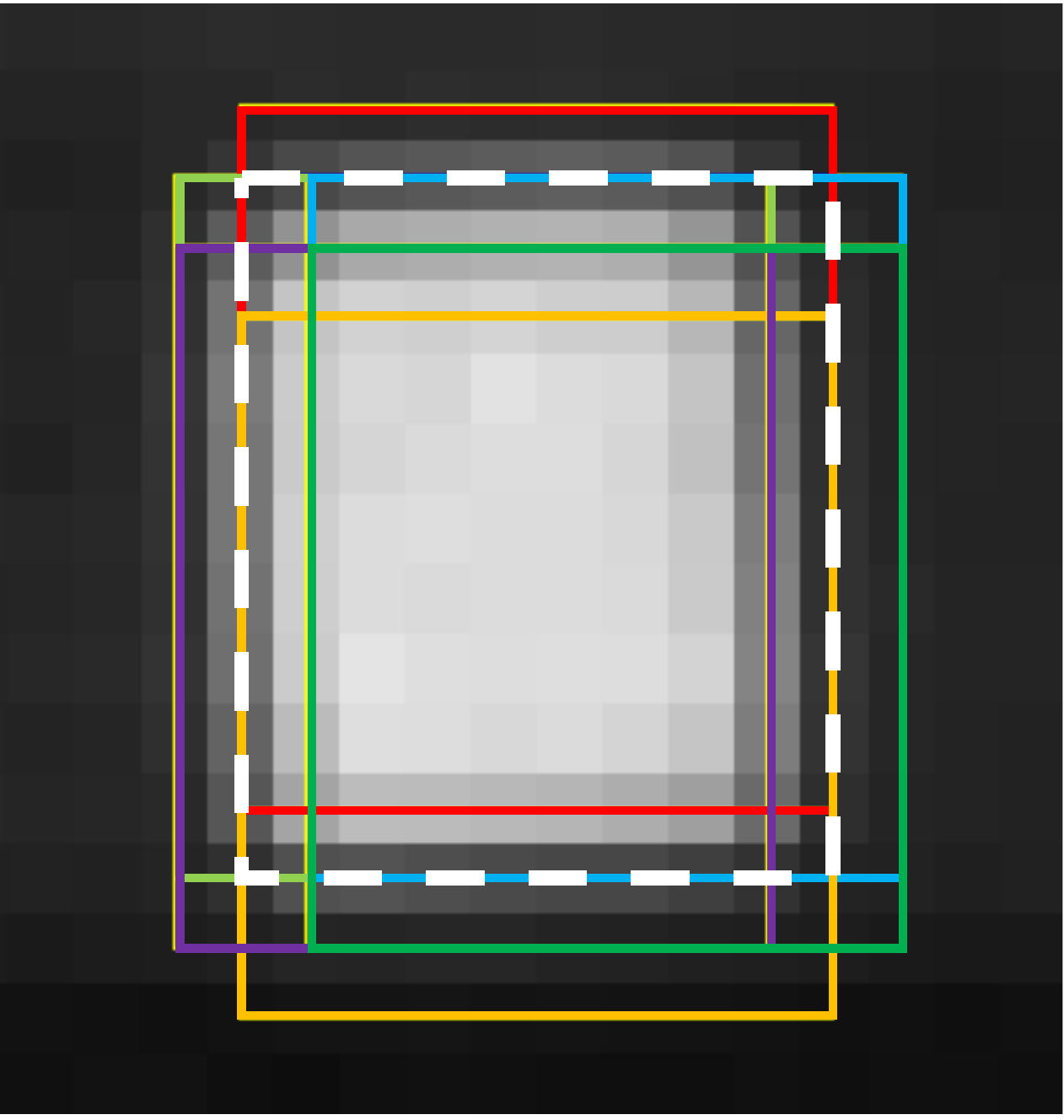}
	\includegraphics[height=0.3\textwidth]{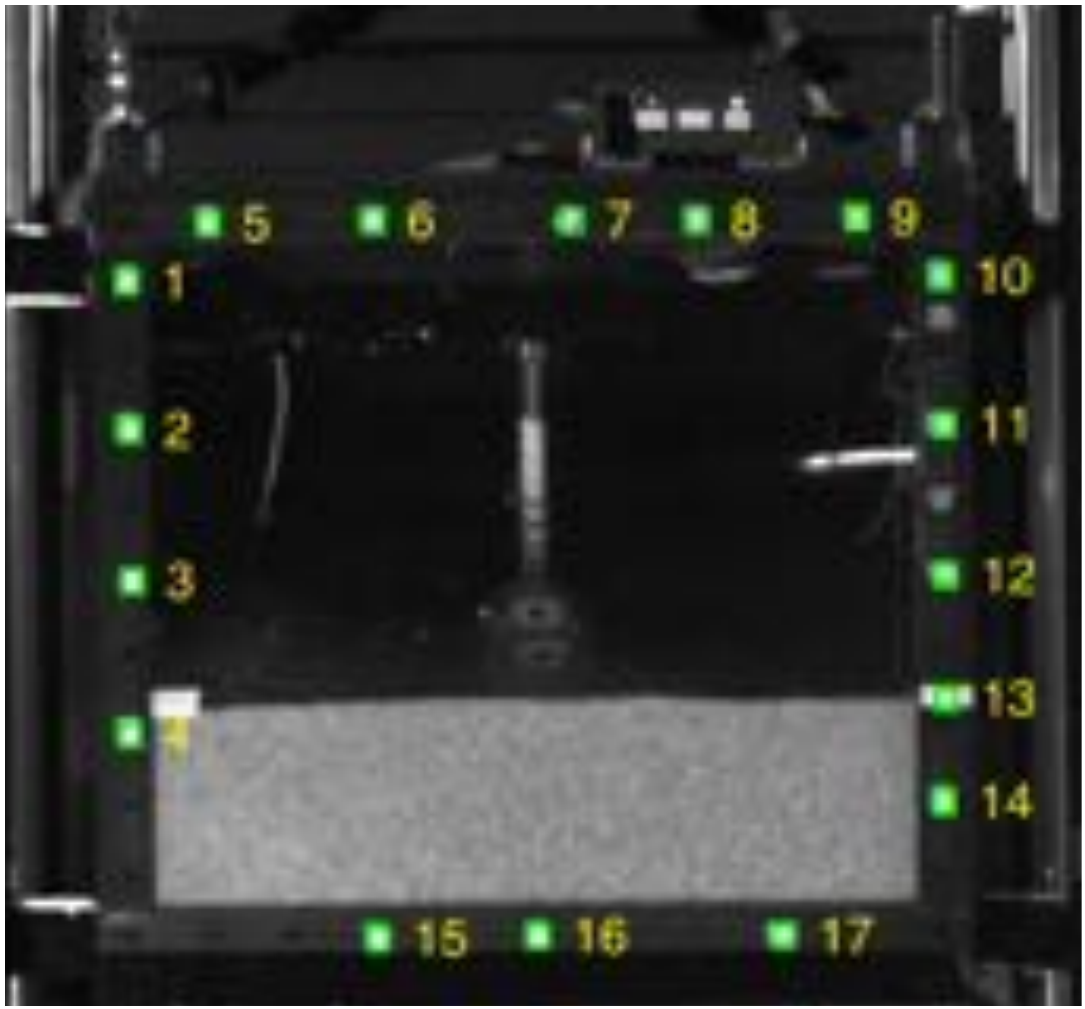}
	\caption{{\bf Reference marker identification.}  Left: The search area to find the multiple reference markers of the sand container is shown in yellow. Middle: For each reference marker several candidates are found (shown by the different coloured lines). The option with the closest match to the reference model matrix is selected (indicated by the white dashed line). Right:  The reference markers identified in the image.}
	\label{fig:markers}
\end{figure*}

Once the coordinates of the sand container have been automatically obtained from the previous step, a search area to find the multiple reference markers of the sand container can be defined as shown in \fig{markers} (left). The model reference marker is a rectangular white matrix. As for the sand container search algorithm, the search area is divided into sub-matrices with the same dimensions as the model matrix and the MLSE is calculated between the model matrix and each sub-matrix. From this search several candidate reference markers appear as each reference marker is identified several times (see \fig{markers}, middle). The best fit for each reference marker amongst the candidates is selected by grouping the candidates according to their coordinates and selecting the option with the minimum error (represented by the white dashed line in \fig{markers}, middle). Using this method, and the dynamical model to reduce the search area as described above, the reference markers are identified in each image (see \fig{markers}, right).

\subsubsection{Motion estimation }

\begin{figure*}
	\centering
	\includegraphics[width=0.5\textwidth]{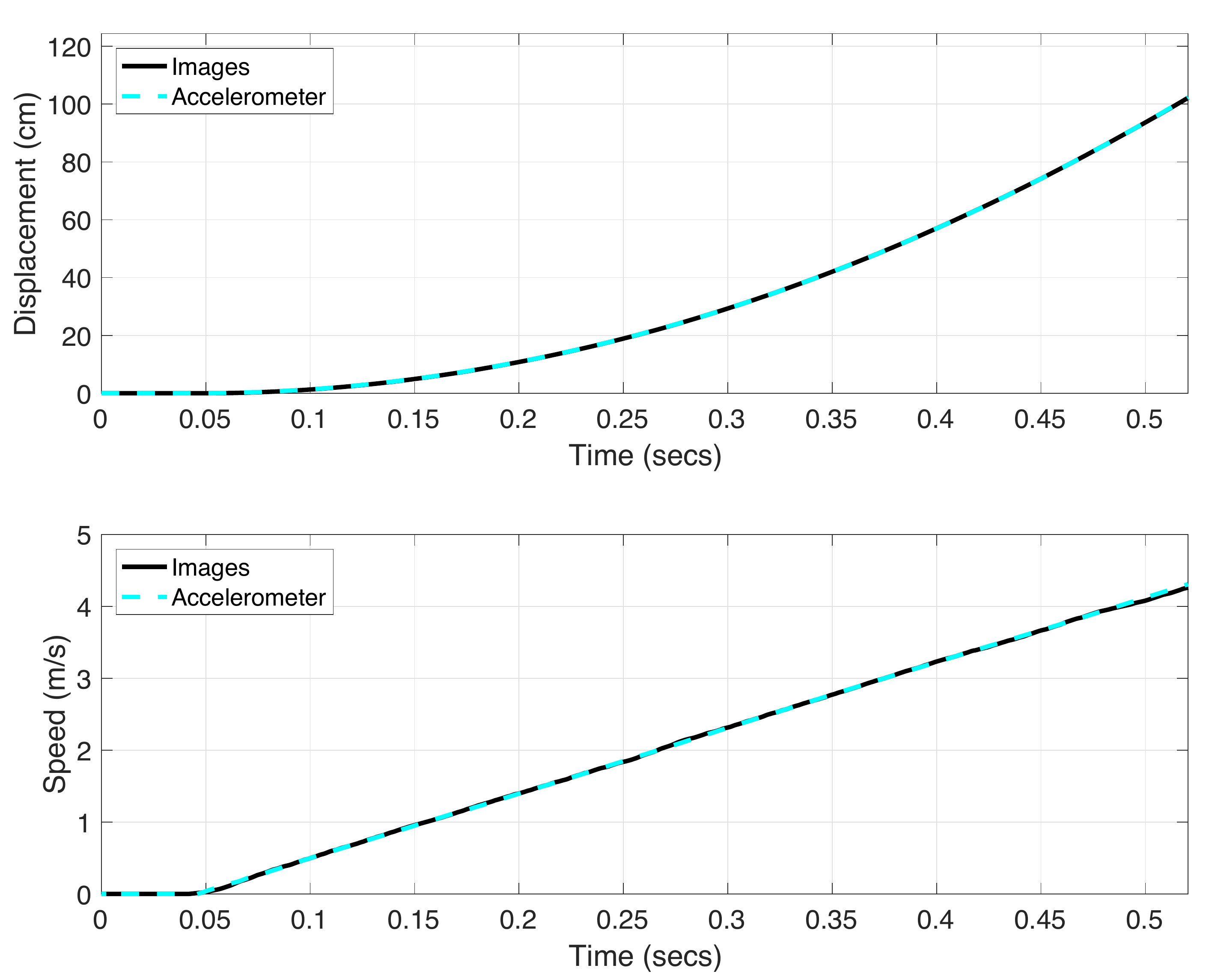}
	\caption{{\bf Comparison of image and accelerometer motion estimation of the sand container.} The displacement (above) and velocity (below) of the sand container as a function of time as determined from the image analysis (black) and accelerometer data (cyan). The image analysis provides the container displacement directly and the container velocity is the derivative of the displacement. The `accelerometer' velocity and displacement are determined by integrating the accelerometer data. In this trial the sand container is dropped from a height of 1.8 m and there are 4.8 kg of counterweights attached to the pulleys, including the mass of the counterweight holders.   }
	\label{fig:ImagesAccComp}
\end{figure*}

Once the markers have been identified in every image, the coordinates of the top left pixel of each reference marker are recorded allowing the displacement in pixels to be calculated.  The displacement of the sand container is then the mean displacement of all of the reference makers and the velocity is the time derivative of the mean displacement. By measuring the sand container width in the images, the pixel scale was determined to be 5 pixels per cm. \Fig{ImagesAccComp} compares the displacement and velocity of the sand container as a function of time, calculated from the image and accelerometer analysis. The two independent methods show a good agreement for both the velocity and the displacement. However, as the process of numerical differentiation introduces more errors than integration, the displacement determined from the accelerometer data is more accurate than the acceleration determined from the image-based displacement data. In fact, the acceleration data is not shown here because small fluctuations in the image-based displacement and velocity result in large fluctuations after differentiation. The accelerometer data, therefore, will be used for the remaining analysis.

\section{Results}
\begin{figure*}
	\centering
	\includegraphics[width=0.48\textwidth]{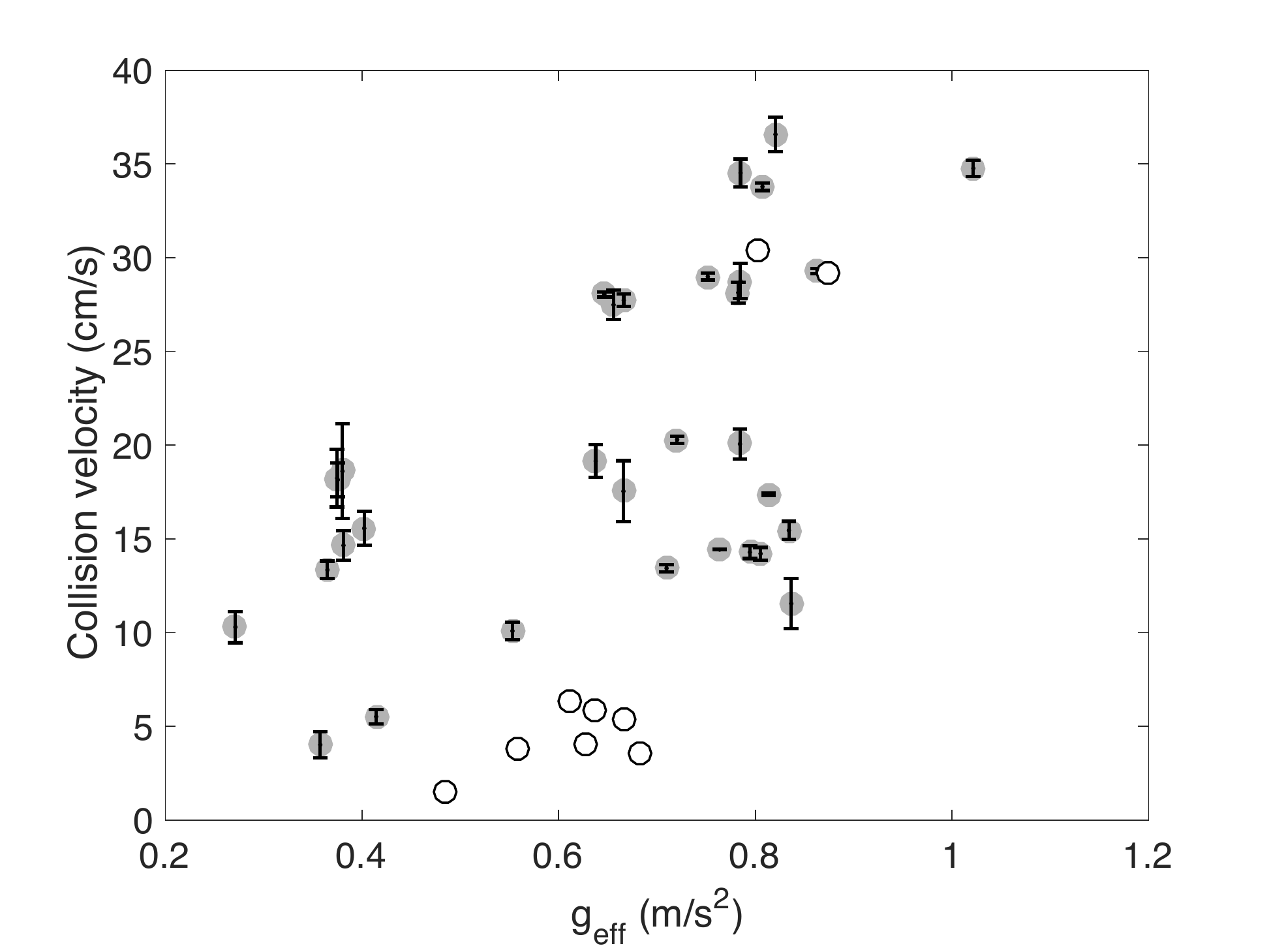} 
 \includegraphics[width=0.48\textwidth]{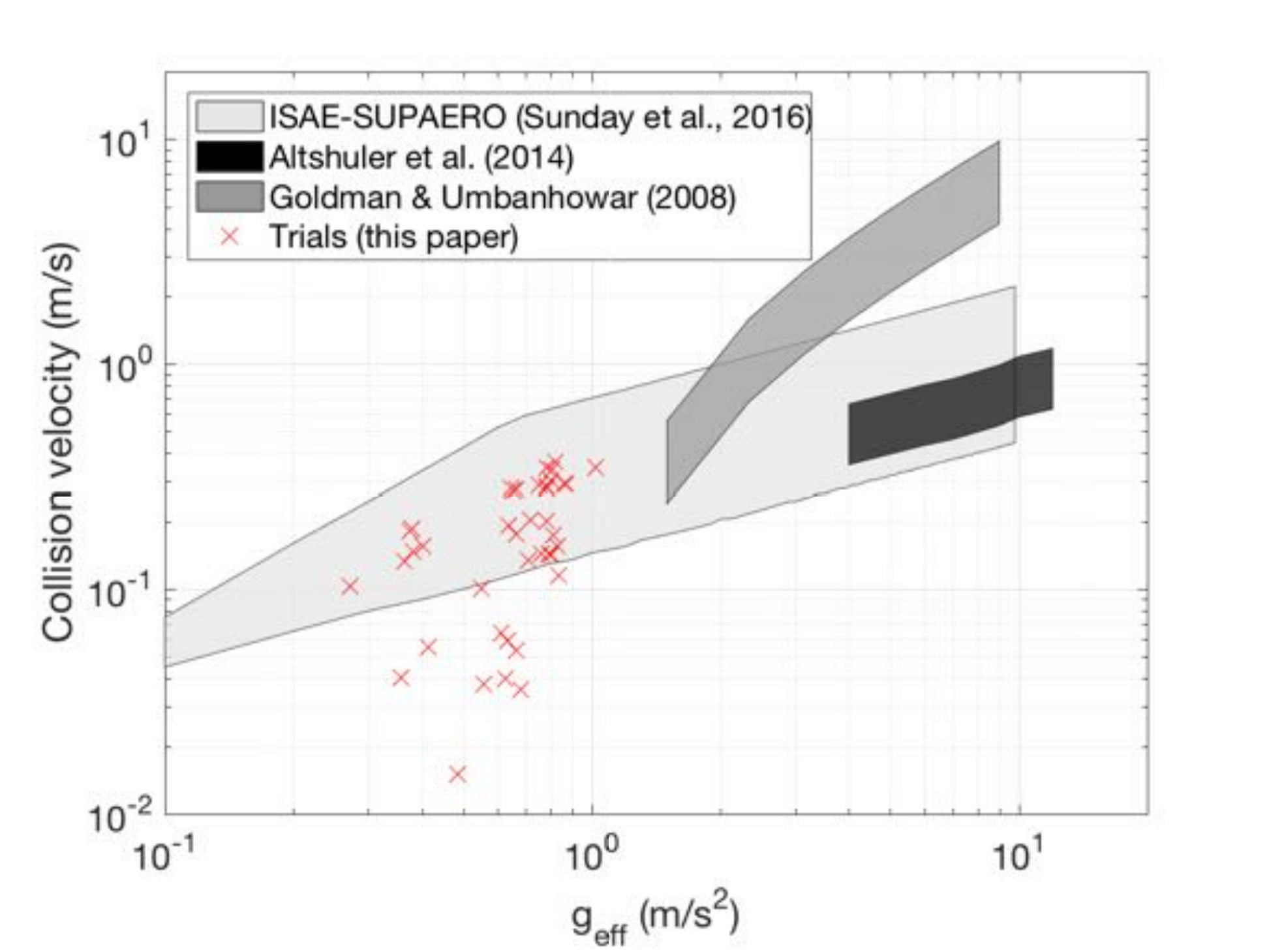} 
	\caption{{\bf Experiment Range. } Left: The range of the experimental trials performed. At low effective gravities and high collision velocities, the experiment is limited by the experiment height. At high effective gravities and low collision velocities, the experiment is limited by the initial separation between the projectile and the sand. See \protect\cite{sunday2016} for details.  The hollow markers show trials were there was only one functioning accelerometer in the projectile. The solid markers show trials where there were two functioning accelerometers in the projectile, and the error bars indicate the standard deviation of the two measurements.  Error bars are not shown for the effective acceleration as there was only one accelerometer attached to the surface container for these trials. Right: Relative regimes accessible from known Atwood machines \protect\citep[for details see][]{sunday2016}. Also shown (red crosses) are the trials that have been performed here.}
	\label{fig:range}
\end{figure*}

\noindent 
\Fig{range} gives the range of experiments performed.  The range of experiments performed is limited due to the experimental set-up; at low effective accelerations and high collision velocities, the experiment drop height is the limiting factor. At low collision velocities and higher effective accelerations, the initial separation between the projectile and the sand surface is the limiting factor. This is discussed in detail in \cite{sunday2016}. The range of the performed trials, compared to the ranges of other Atwood machine collision experiments and the theoretical calculations of the ISAE-SUPAERO drop tower (presented in \cite{sunday2016}) is given in \Fig{range}, right. The main reason for the differences between the theoretical experiment range and the actual experiment range is the friction in the drop tower guide rails; as the sand container accelerates slightly slower than in the case where there is no friction, the projectile - sand collision occurs sooner and the relative velocity is smaller. As the in-situ effective acceleration is measured directly by the surface container accelerometer, the variation from the theoretical results only has consequences for the experiment planning (choice of counterweight masses, choice of drop height to ensure that the collision is centered in the camera field of view, ...) but does not affect the scientific results. The measured range of the trials confirms that we have developed a robust experimental method for performing low-velocity collisions in reduced-gravity. 

\begin{figure*}
	\centering
\includegraphics[width=0.45\textwidth]{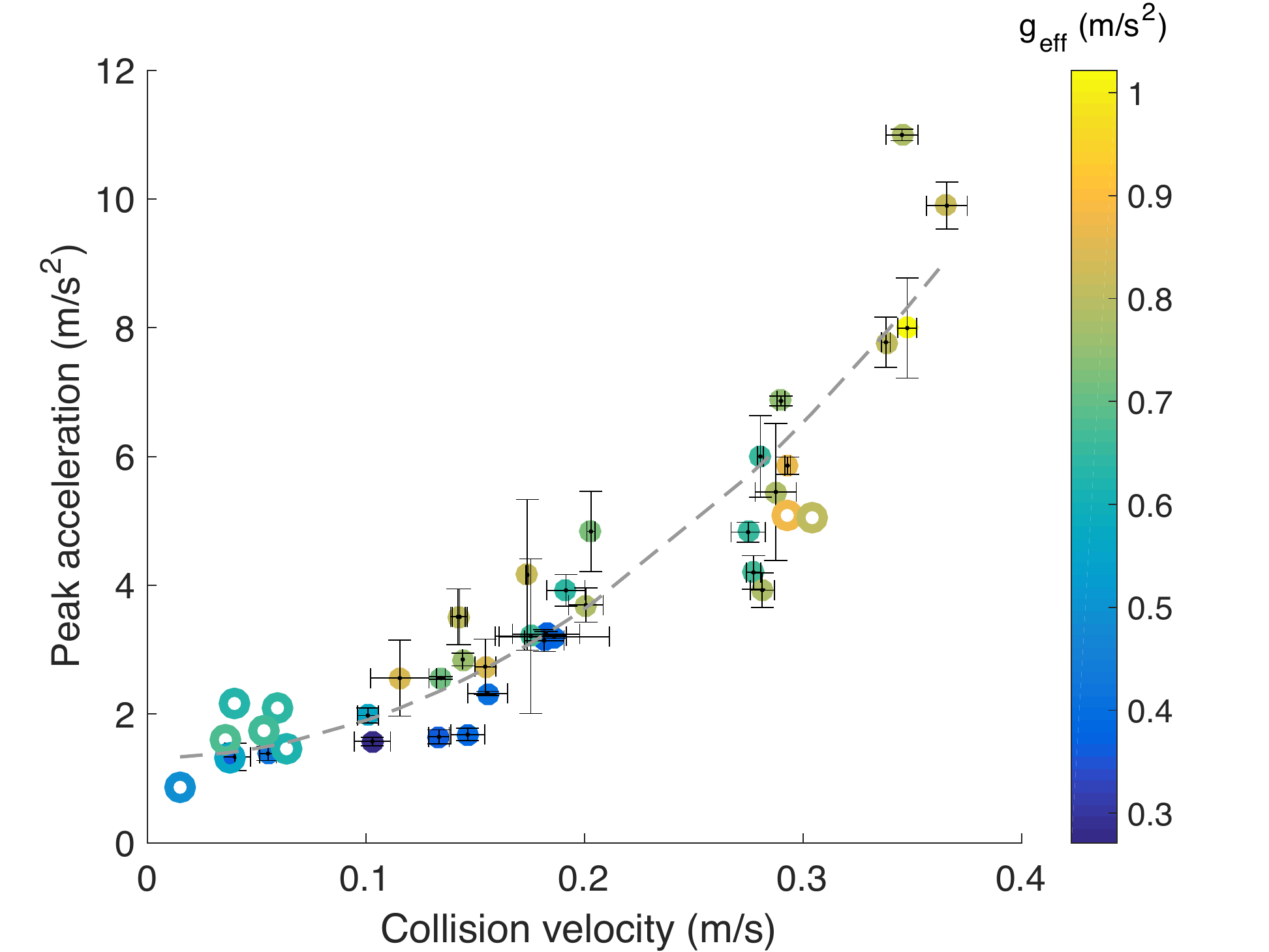} 
	\includegraphics[width=0.45\textwidth]{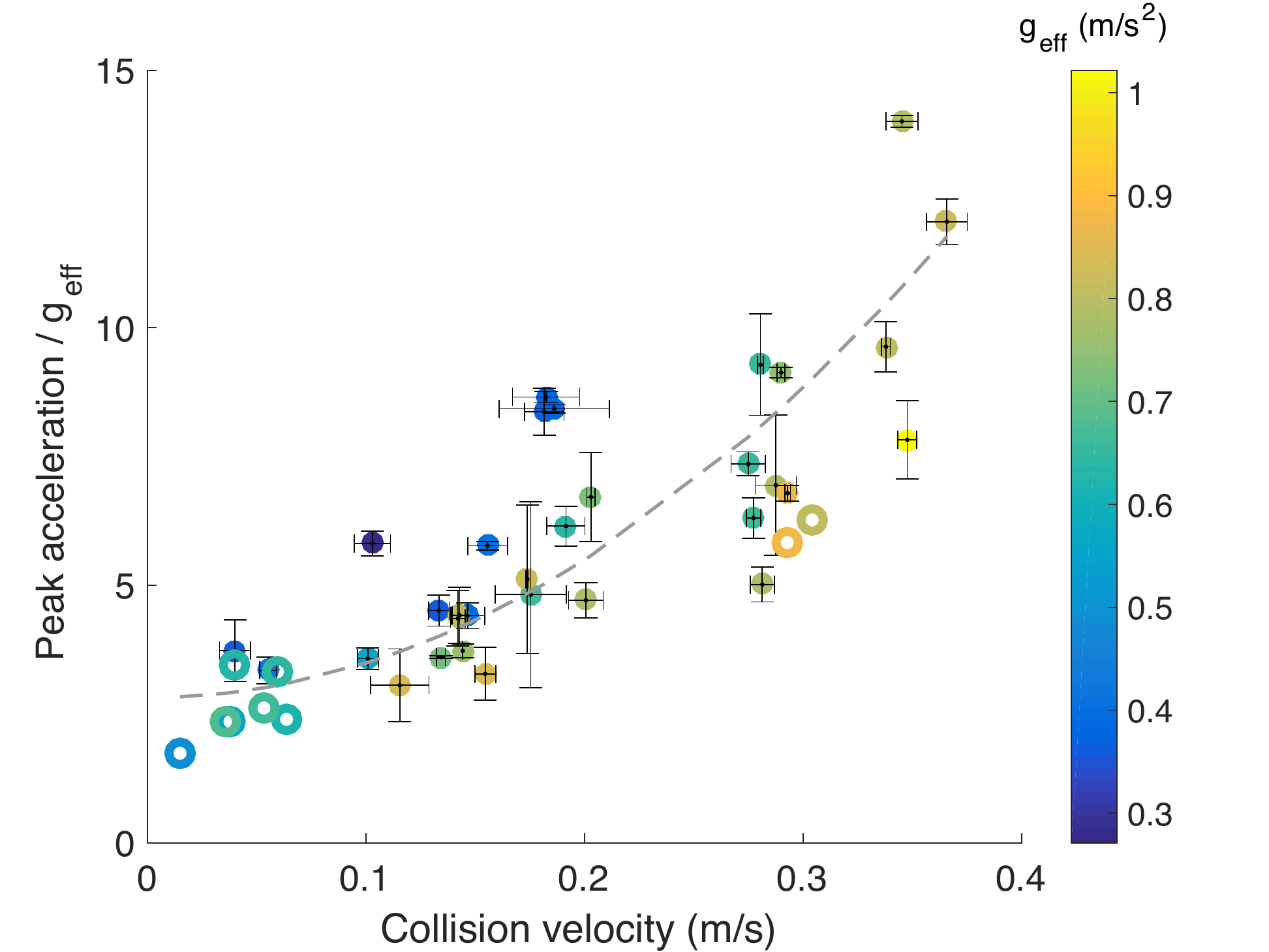} 
	\caption{{\bf Peak acceleration and collision velocity. } Left: Peak acceleration of the projectile as a function of the collision velocity. Right: Peak acceleration of the projectile normalised by the effective gravity as a function of the collision velocity. The markers are colour-coded to indicate the effective gravity, as shown in the colour bar. The solid and hollow markers have the same significance as in \fig{range}. The dashed grey lines in the figures show the following fits to the data: $A_{max} = 57.9 V_c^2  + 1.3$ and  $A_{max}/g_{eff} = 66.9 V_c^2 + 2.8$, respectively.}
	\label{fig:PeakAccVelocity}
\end{figure*}

The peak acceleration measured by the projectile sensors during the collision with the sand surface has the strongest dependence on the collision velocity and scales with the square of the collision velocity (\fig{PeakAccVelocity}).  At the lower collision velocities, the peak is not pronounced, but is rather a broad maximum \citep[as also observed by][]{goldman08}. The trend of increasing peak acceleration with collision velocity is clearly visible in both the unnormalised and normalised peak acceleration data (\fig{PeakAccVelocity}).  The peak acceleration does not tend to zero as the collision velocity approaches 0 cm/s. This non-zero intercept indicates a force dominated by friction; frictional/hydrostatic forces dominate at very low collision velocities, compared to higher velocities where hydrodynamic forces dominate (see \sect{discussion}).

\begin{figure*}
	\centering
	\includegraphics[width=0.45\textwidth]{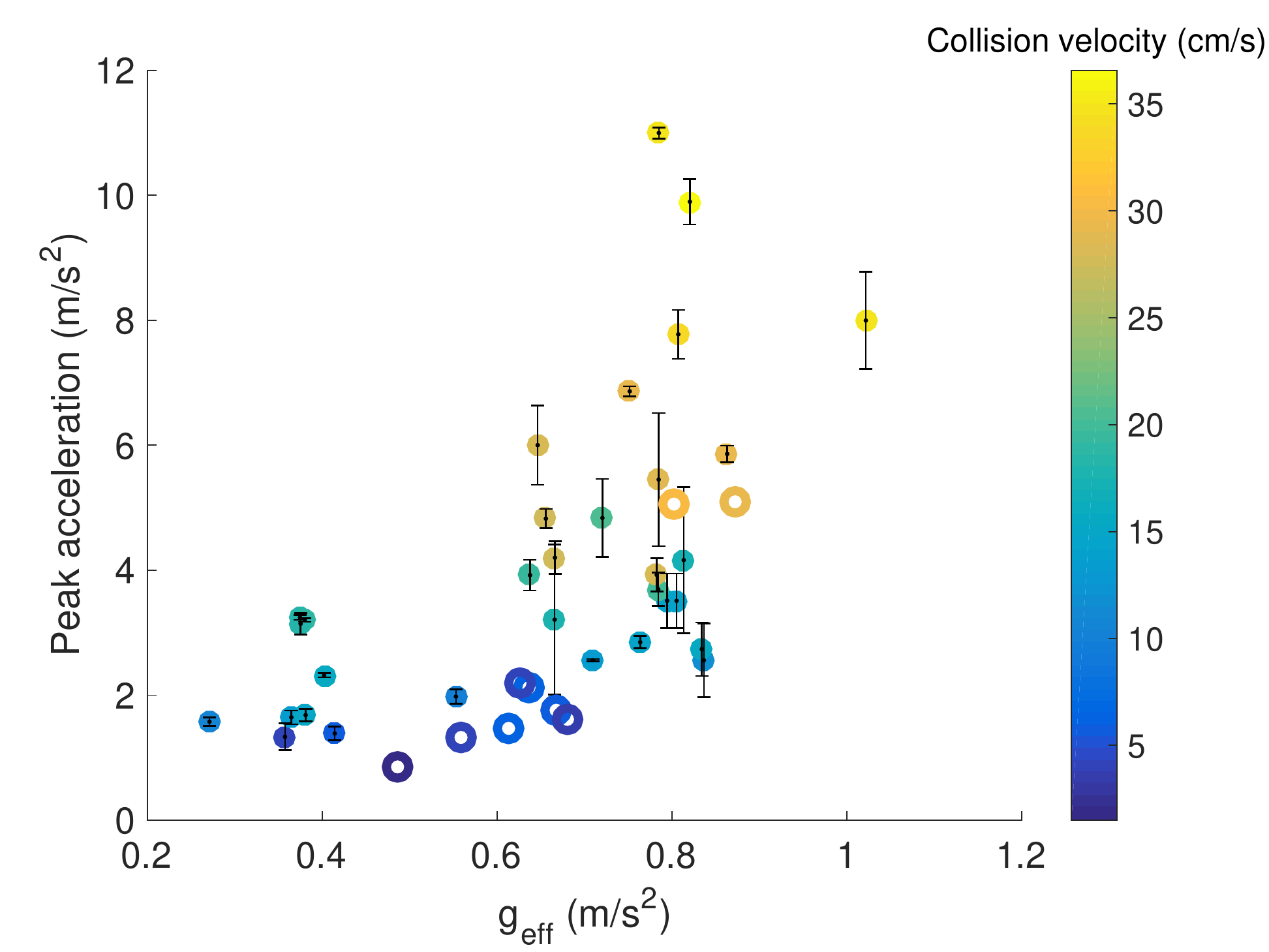} 
	\includegraphics[width=0.45\textwidth]{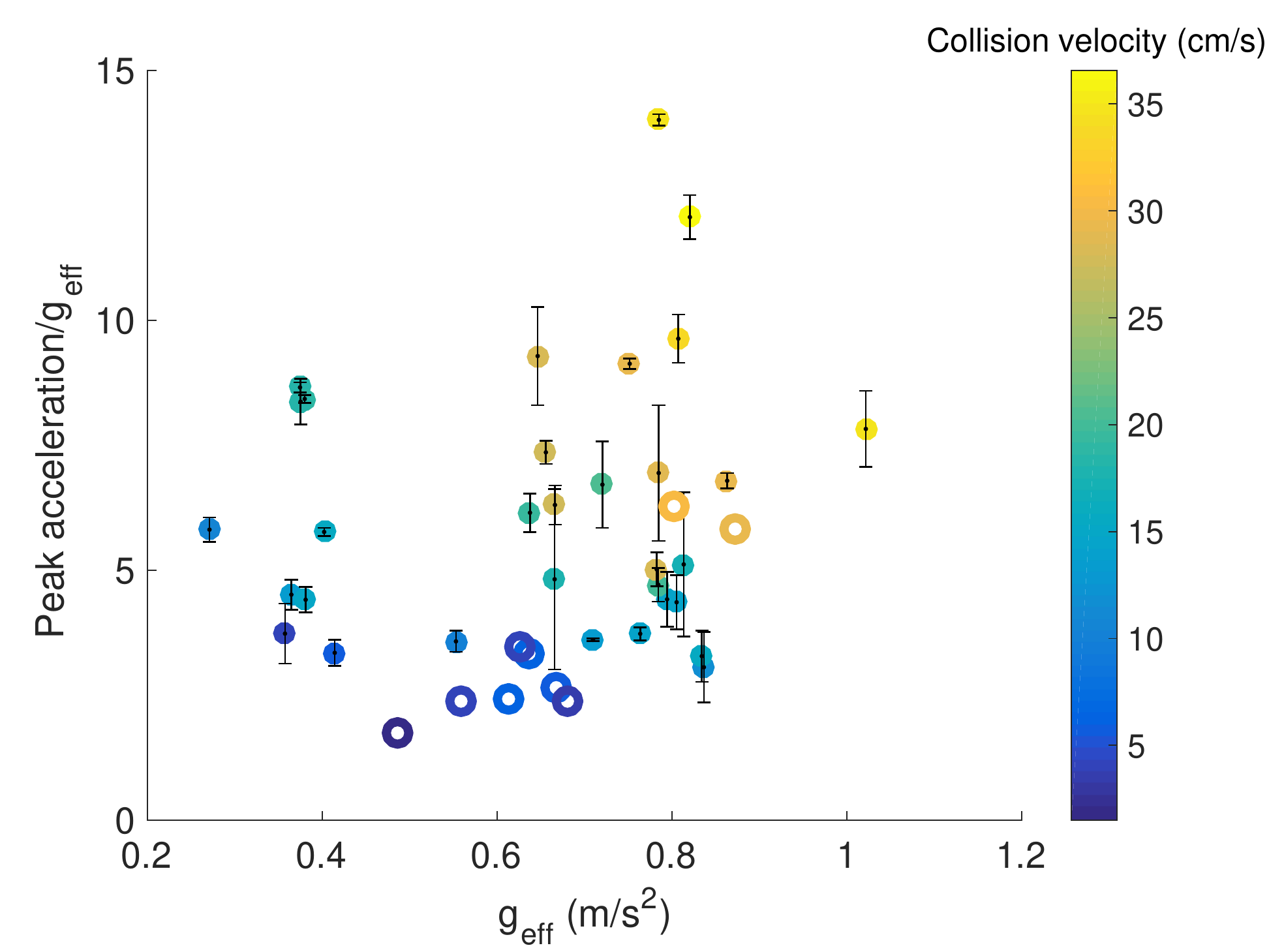} 
	\caption{{\bf Peak acceleration and effective gravity. } Left: Peak acceleration of the projectile as a function of the effective gravity (the measured acceleration of the surface container). Right: Peak acceleration of the projectile normalised by the effective gravity as a function of the effective gravity.  The markers are colour-coded to indicate the collision velocity, as shown in the colour bar. The solid and hollow markers have the same significance as in \fig{range}.}
	\label{fig:PeakAccGeff}
\end{figure*}

 \Fig{PeakAccGeff} shows the peak acceleration measured by the projectile sensors during the collision with the sand surface, as a function of the effective gravity \ie the measured acceleration of the surface container, for all of the experimental trials performed.  The apparent trend of increasing peak accelerations with increasing effective gravity can be explained by the range of experimental trials that were performed; a larger number of low-velocity collisions were performed at lower effective gravity levels and vice-versa (\fig{range}). Therefore, as the trials with larger collision velocities have larger peak accelerations (\fig{PeakAccVelocity}) an experimental bias is created in the data. Trials of similar collision velocities with different effective gravity levels actually show similar peak accelerations. 

\begin{figure*}
	\centering
\includegraphics[width=0.45\textwidth]{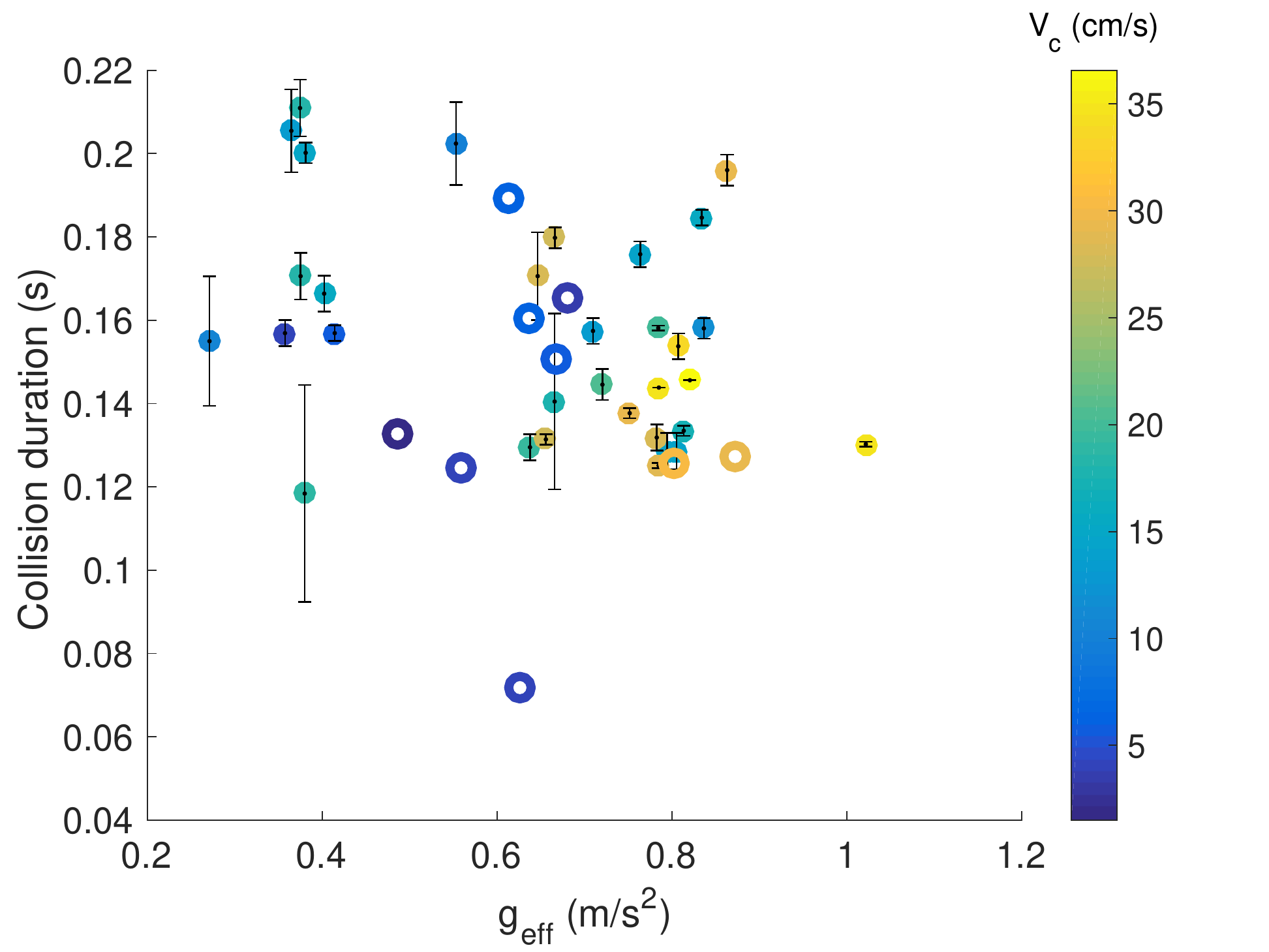} 
	\includegraphics[width=0.45\textwidth]{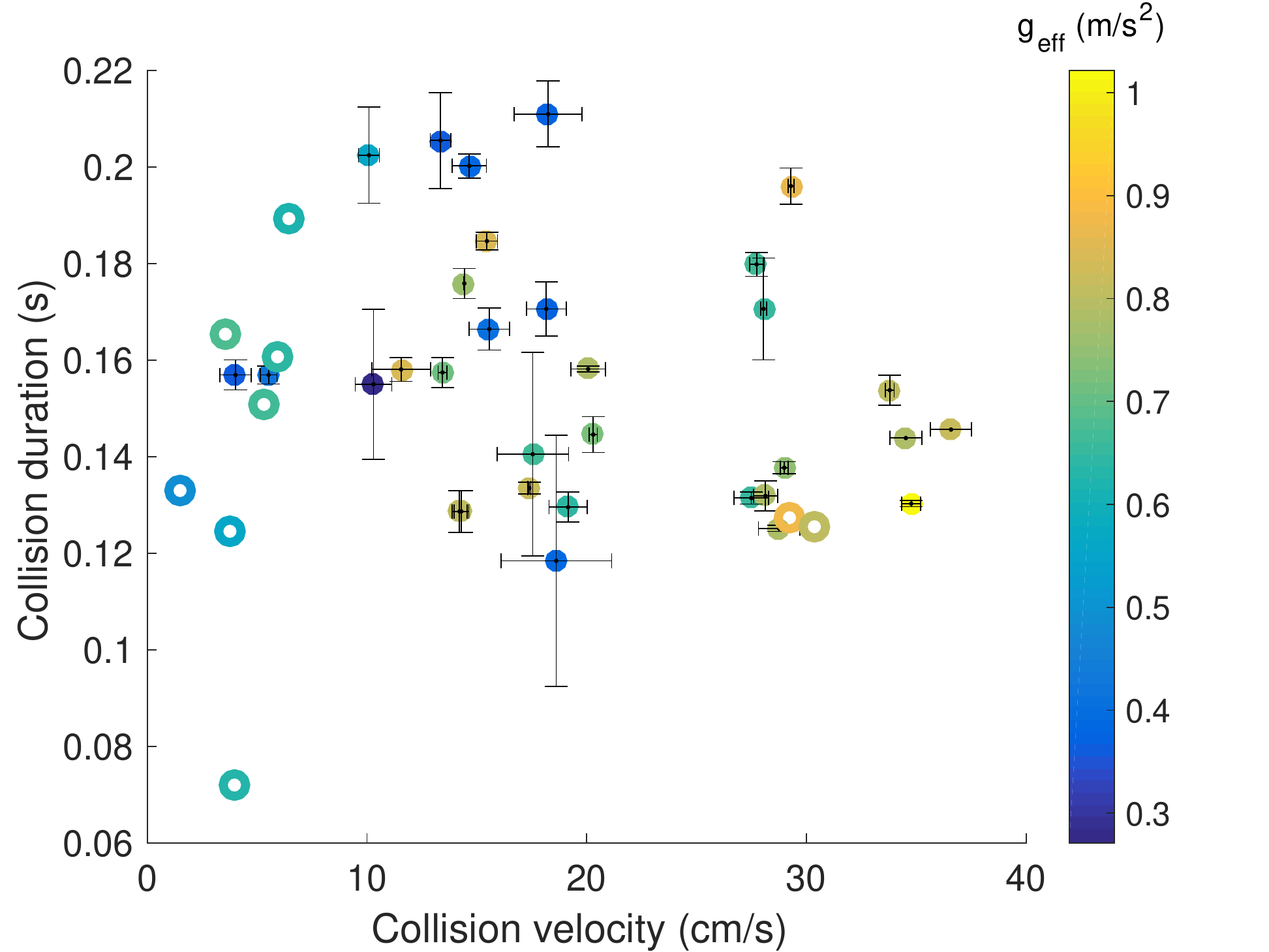} 
	\caption{{\bf Collision duration. }Left: Collision duration as a function of the effective gravity. The markers are colour-coded to indicate the collision velocity, as shown in the colour bar.  Right: Peak accelerations as a function of the collision velocity. The markers are colour-coded to indicate the effective gravity, as shown in the colour bar. The solid and hollow markers have the same significance as in \fig{range}.}
	\label{fig:Duration}
\end{figure*}

\begin{figure*}
	\centering
\includegraphics[width=0.45\textwidth]{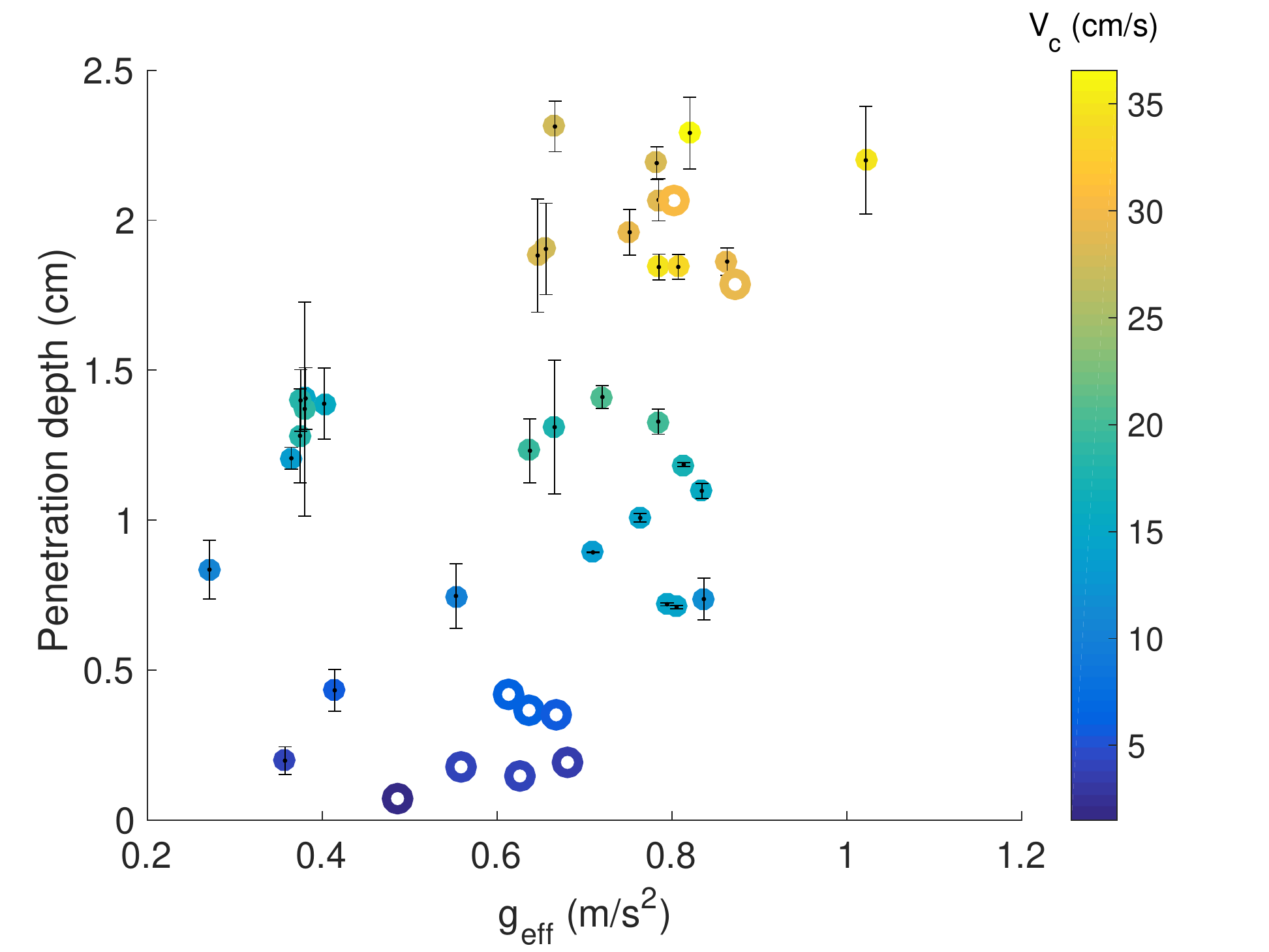} 
	\includegraphics[width=0.45\textwidth]{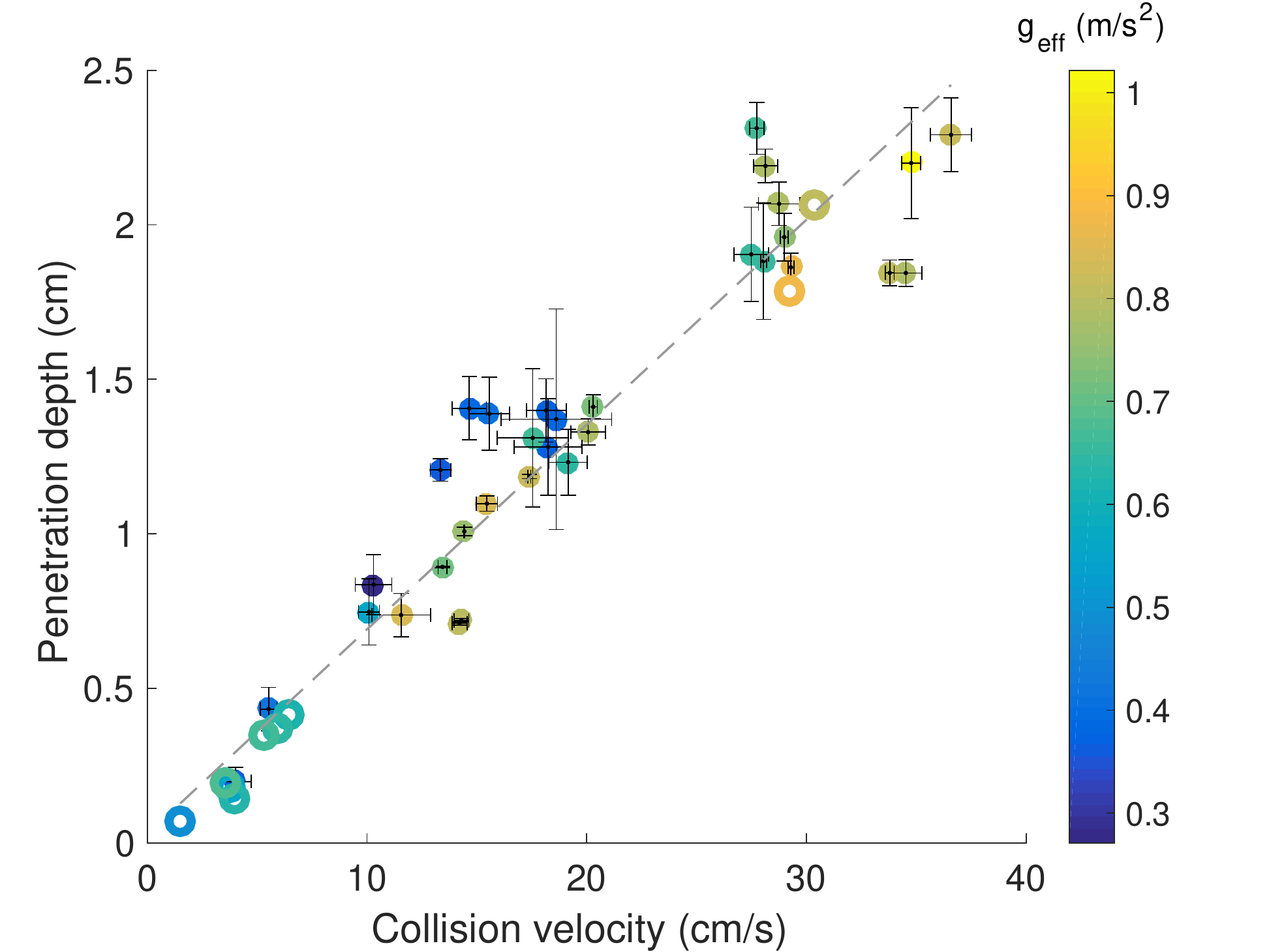} 
	\caption{{\bf Penetration depth. }Left: Maximum penetration depth as a function of the effective gravity. The markers are colour-coded to indicate the collision velocity, as shown in the colour bar.  Right: Maximum penetration depth as a function of the collision velocity. The markers are colour-coded to indicate the effective gravity, as shown in the colour bar. The dashed grey line shows the best linear fit to the data:  $Z_{max} =  0.07 V_c + 0.03$.  The solid and hollow markers have the same significance as in \fig{range}.}
	\label{fig:Depth}
\end{figure*}

The collision duration ranges from $\sim$70 ms to $\sim$210 ms but is independent of both the effective gravity and the collision velocity (\fig{Duration}). Our data indicate that the maximum penetration depth is also independent of the effective gravity but scales linearly with the collision velocity\footnote{To verify the linear dependence, a generic model $Z_{max} = \alpha V_c^n + \beta$ has been tested with 200 values between 0 and 2. The minimum least squares error corresponds exactly to $n$ = 1.} (\fig{Depth}).  This dependance of penetration depth on collision velocity is expected given that the deceleration of the projectile scales with the collision velocity (\fig{PeakAccGeff}). 

The acceleration profile of the projectile always reaches a maximum (for lower velocities, the maximum can be very broad) before reducing to the level of the surface container (see \fig{typical} for an example).  As the projectile and the surface container have the same acceleration at the end of the experiment, this indicates that the collision has entirely finished within the period of time of the drop. Also, as the projectile acceleration does not return to 0 m/s$^2$, the projectile does not return to free-fall and has, therefore, not left contact with the sand surface. In all of the experimental trials performed here, no rebound was observed and the coefficient of restitution is, therefore, zero.

\section{Discussion}
\label{s:discussion}
\noindent  The drag force during penetration into a granular can be separated into two terms:  the hydrodynamic drag force, and the static resistance force \citep[\eg][]{Allen57}.  If the drag force scales with the square of the collision velocity, this indicates that the impact is occurring in the inertial (or hydrodynamic) regime. In other words, that the grains have become sufficiently fluidised during the impact for the system to display inertial, fluid-like drag.  On the other hand, a small or zero dependance of the drag force on the collision velocity would indicate a quasi-static (or hydrostatic) regime. 
 
In our experiments, the peak accelerations (and thus the drag force) scale as $A_{max} = 57.9 V_c^2  + 1.3$, where the units of $A_{max}$ and $V_c$ are m/s$^2$ and m/s, respectively (see \fig{PeakAccVelocity}). Therefore, for collision velocities of $\gtrsim$15 cm/s, the hydrodynamic drag dominates and the collisions occur in the inertial regime, and for collision velocities of $\lesssim$15 cm/s, the static resistance force dominates and the collisions occur in the quasi-static regime. For the regime transition to occur at such low velocities is a surprising result.  \cite{goldman08}, for example, found that under terrestrial gravity the quasi-static to inertial regime change occurs at collision velocities an order of magnitude larger: the peak acceleration of two steel spheres ($\sim$2 kg and $\sim$80 grammes) impacting glass beads scales with the square of the collision velocity for velocities $\gtrsim$1.5 m/s, but not for the lower collision velocities.  In the same experiments, this regime change was also evident in the collision durations: the collision time is independent of the collision velocity at the higher collision velocities (in the inertial regime), but for lower collision velocities ($\lesssim$1.5 m/s), the collision duration increases with decreasing collision velocity indicating the quasi-static regime.  Here we find that the collision duration is independent of the collision velocity \citep[such as in the higher energy collisions of ][]{goldman08}, further indicating the importance of the inertial regime in our experiments.

One explanation for the regime change occurring at collision velocities an order of magnitude smaller in our experiments (and impact energies two orders of magnitude smaller) is the different surface materials used. In our experiments, the quartz sand is more irregular and frictional that the glass beads used by \cite{goldman08} \citep[yet still much less angular than the samples returned from asteroid Itokawa;][]{tsuchiyama2011}. The hydrodynamic drag will, therefore, be of greater importance here than in their experiments. The same applies if we compare our results to those of \cite{altshuler14} who studied a sphere impacting polystyrene beads.  Given the impact velocities of $\sim$1 m/s and the small hydrodynamic drag coefficient of polystyrene beads, they estimate the hydrodynamic drag in their experiments and simulations to be negligible. The different surface materials may, therefore, also explain why we find the collision duration to be independent of the effective gravity whereas both \cite{altshuler14} and \citep{goldman08} found that the collision duration scales with $g_{eff}^{1/2}$. 

An alternative, or perhaps additional, explanation can be found by recalling that the quasi-static resistance force is proportional to the object's cross-section times the local pressure \citep{albert99}. As the local pressure is directly related to the gravitational acceleration, the quasi-static resistance force should tend to zero as gravity is reduced.  \cite{katsuragi07} also suggest that the quasi-static force is linearly proportional to gravity in granular impact cratering experiments.  Therefore, in the absence of gravity or a confining pressure, the quasi-static regime does not exist and the drag force should scale with the square of the velocity for any velocity range, not just for high-energy impacts.  This was also observed by \cite{seguin2016} who performed numerical simulations of a sphere moving through a cloud of grains. They explain that, where no gravity acts and no external pressure is imposed from any external boundary, no stress scale exists except the kinetic pressure ($\rho V^2$) arising from the collision processes. The quasi-static regime, therefore, is expected to reduce as the effective gravity becomes lower. We indeed observe that the quasi-static to inertial regime transition occurs for much lower impact energies in our low-gravity experiments compared to similar experiments performed under terrestrial gravity. 

Our data indicate that the maximum penetration depth is independent of the effective gravity \citep[as also found by][]{altshuler14} but scales linearly with the collision velocity.  The linear scaling of penetration depth with impact velocity was also found by  \cite{goldman08} and \cite{Bruyn2004} for higher velocity ($>$1 m/s) impact experiments in which the penetration depth is generally greater than one projectile radii. We show here that, at low effective gravities, the linear scaling is also valid for shallow impacts of an aluminium sphere into sand. The projectile has a diameter of 10 cm and the maximum penetration depth observed in the experiments is $\sim$1/4 of the projectile diameter. 

\section{Conclusions}
\noindent  Making use of our novel drop tower facility \citep{sunday2016}, we have performed low-velocity (2 - 40 cm/s), shallow impact experiments of a 10 cm diameter aluminum sphere into quartz sand in low effective gravities ($\sim0.2 - 1$ m/s$^2$).  A total of 46 trials were performed, of which 41 were classified as acceptable trails and subsequently analysed. No rebounds are observed in the experimental trials and the coefficient of restitution is thus zero. We find that the penetration depth scales linearly with the collision velocity but is independent of the effective gravity for the experimental range tested, and that the collision duration is independent of both the collision velocity and the effective gravity. 

During similar impact experiments under terrestrial gravity, the transition from the quasi-static regime (where the static resistance force dominates) to the inertial regime (where hydrodynamic effects dominate) occurs at collision velocities of $\sim$1.5 m/s \citep{goldman08}.  Our low-gravity experimental results indicate that the collisions occur in the inertial regime down to collision velocities of $\sim$15 cm/s; equivalent to an impact energy two orders of magnitude smaller than the terrestrial gravity experiments. The lower energy regime change may be due to the increased hydrodynamic drag in our experiments, but may also support the notion that the quasi-static regime is expected to reduce as the effective gravity becomes lower. The latter indicates that, in the absence of gravity or a confining pressure, the quasi-static regime does not exist and the drag force scales with the square of the velocity for any velocity range, not just for high-energy impacts. 

To investigate the reason(s) for the lower energy regime change found here further low-gravity experiments should be performed with different surface materials. It would also be useful to perform experiments at similar collision velocities over a large range of effective gravities; a challenging task given the range of experimental conditions accessible in our drop tower (\fig{range}). Numerical simulations \citep[\eg][]{schwartz2012,sanchez2011,holsapple1993}, validated using the experimental data obtained in these trials, may be a complementary approach for such studies. The numerical simulations could also be used to extrapolate our results to even lower gravity regimes, inaccessible with our drop tower. For comparison, the smallest effective gravity obtained in these experiments is just less than that of asteroids (1) Ceres and (4) Vesta, with surface gravities of $\sim$0.29 m/s\textsuperscript{2} and $\sim$0.25 m/s\textsuperscript{2}, respectively. The $\sim$17 km asteroid (433) Eros has a surface gravity one hundred times smaller than the smallest effective gravity tested here, and the surface gravity of (25143) Itokawa ($\sim$300 m) is yet another hundred times smaller. On the other end of the scale, the largest effective acceleration that has been tested in these experiments ($\sim$1 m/s\textsuperscript{2}) is comparable to the surface gravity of Saturn's moon Enceladus. 

Finally, in order to improve the data analysis, it may be possible to combine both the images and the accelerometer data. By using information theory \citep[\eg][]{Khaleghi2013}, the combination of relevant information from these two data sources may provide an even more reliable measure of the motion of both the surface container and the projectile.

\section{Implications for small body space missions and for asteroid evolution}

\noindent  Current and future asteroid landers \citep[MASCOT, MASCOT-2, MINERVA, AGEX; ][]{tsuda2013,Ho2016,karatekin2016} will be deployed on ballistic trajectories to the asteroids' surfaces with no attitude control. If during their low-velocity landing there is no rebound (as was the case in these experiments), the lander would remain at the location of the initial touchdown. This places even more importance on the precision of the deployment strategy from the main spacecraft for the landers without mobility mechanisms. However, this would also reduce the risk of the lander rebounding and being lost to space, and would simplify the lander design as specific areas of the asteroid surface could be targeted \citep[\eg choosing a region with the best thermal conditions for the payload operations in order to minimise, as far as possible, complex thermal regulation; ][]{Cadu2016EGU}. 

Our experiments suggest that the landing velocity is the critical parameter that will influence both the penetration depth and the acceleration profile during landing.  The peak accelerations observed varied from approximately 1 to 12 m/s$^2$ and the maximum penetration depth observed in the experiments was $\sim$1/4 of the projectile diameter. However, a harder, or denser, surface material is likely to lead to larger peak accelerations and a smaller penetration depth, whereas a more fluffy regolith may reduce the peak accelerations while increasing the penetration depth.  The variation of peak accelerations and penetration depth with surface properties (for example particle size, density, cohesion, angularity and frictional properties) in low-gravity should be studied in future experiments in order to cover as many asteroid surface materials as possible. Similarly, for shallow penetrations, the projectile shape is known to play crucial role, with sharper objects penetrating deeper \citep{newhall2003}.  Further experiments are, therefore, needed to understand the influence of the projectile shape in low-velocity and low-gravity collisions.  This is also particularly important since asteroid landers are often rectangular rather than spherical in shape \citep[\eg][]{Ho2016,karatekin2016}.  

The observed penetration rather than rebounding also has implications for asteroid surfaces. As discussed in \cite{Nakamura2008}, two types of  low-velocity  impact  can  occur naturally on asteroids: the impact of remnant small ejecta from a catastrophic disruption event, and the secondary impact of ejecta blocks from a primary impact on the surface.  The collision dynamics between the impacting ejecta and the regolith on an asteroid surface determines whether the debris will bounce, penetrate fully or partially into the surface, or remain on the surface. Our results support the findings of \cite{Nakamura2013} showing that the isolated large blocks on the smooth terrains of Itokawa \citep[\eg][]{Nakamura2008} could be ejecta blocks that collided with the surface during re-accumulation, but did not deeply penetrate. If the inertial regime dominates on asteroid surfaces, the penetration depth of ejecta from cratering events will also be shallow, and will be directly linked to the re-impacting velocity of the ejecta. 

More extensive fluidisation has been observed in micro-gravity following changes in the granular force contact network, compared with identical experiments under terrestrial gravity \citep{murdoch13_MNRAS}, and avalanches have been found to be longer range at lower gravity \citep{kleinhans11}. The notion of a reduced quasi-static regime in very low gravity would support the enhanced fluidisation observed in these experiments. This may imply that regolith material becomes more easily fluidised for lower-energy events such as small micrometeoroid impacts \citep[\eg][]{richardson05,garcia15}.  We emphasise, however, that the results presented here are for one experimental configuration only and should be developed further (as mentioned above) to determine more precisely the role of the surface material properties, the impactor properties, and gravity.

\section*{Acknowledgements}

\noindent  This project benefited from some financial support from the Centre National d'Etudes Spatiales (CNES) and from the European Space Agency (ESA). We thank Daniel Gagneux and Thierry Faure, from the D{\'e}partement M{\'e}canique des Structures et Mat{\'e}riaux (DMSM) at ISAE-SUPAERO for their help in designing and building the drop tower experiment, Claudia Valeria Nardi and Sara Morales Serrano for their help in performing the initial drop tower calibration trials, and Tristan Janin and Alexis Calandry for the data collection. Thanks also to Jens Biele for his valuable comments on the manuscript, and to Josh Colwell for his helpful review.





\bibliographystyle{mnras} 
\bibliography{AsteroidsIV,/Users/n.murdoch/Documents/BIBFILE/murdoch.bib} 

\begin{thebibliography}{}
\makeatletter
\relax
\def\mn@urlcharsother{\let\do\@makeother \do\$\do\&\do\#\do\^\do\_\do\%\do\~}
\def\mn@doi{\begingroup\mn@urlcharsother \@ifnextchar [ {\mn@doi@}
  {\mn@doi@[]}}
\def\mn@doi@[#1]#2{\def\@tempa{#1}\ifx\@tempa\@empty \href
  {http://dx.doi.org/#2} {doi:#2}\else \href {http://dx.doi.org/#2} {#1}\fi
  \endgroup}
\def\mn@eprint#1#2{\mn@eprint@#1:#2::\@nil}
\def\mn@eprint@arXiv#1{\href {http://arxiv.org/abs/#1} {{\tt arXiv:#1}}}
\def\mn@eprint@dblp#1{\href {http://dblp.uni-trier.de/rec/bibtex/#1.xml}
  {dblp:#1}}
\def\mn@eprint@#1:#2:#3:#4\@nil{\def\@tempa {#1}\def\@tempb {#2}\def\@tempc
  {#3}\ifx \@tempc \@empty \let \@tempc \@tempb \let \@tempb \@tempa \fi \ifx
  \@tempb \@empty \def\@tempb {arXiv}\fi \@ifundefined
  {mn@eprint@\@tempb}{\@tempb:\@tempc}{\expandafter \expandafter \csname
  mn@eprint@\@tempb\endcsname \expandafter{\@tempc}}}

\bibitem[\protect\citeauthoryear{Albert, Pfeifer, Barab\'asi  \&
  Schiffer}{Albert et~al.}{1999}]{albert99}
Albert R.,  Pfeifer M.~A.,  Barab\'asi A.-L.,   Schiffer P.,  1999, \mn@doi
  [Phys. Rev. Lett.] {10.1103/PhysRevLett.82.205}, 82, 205

\bibitem[\protect\citeauthoryear{Allen, Mayfield  \& Morrison}{Allen
  et~al.}{1957}]{Allen57}
Allen W.~A.,  Mayfield E.~B.,   Morrison H.~L.,  1957, Journal of Applied
  Physics, 28, 370

\bibitem[\protect\citeauthoryear{Altshuler, Torres, González-Pita,
  Sánchez-Colina, Pérez-Penichet, Waitukaitis  \& Hidalgo}{Altshuler
  et~al.}{2014}]{altshuler14}
Altshuler E.,  Torres H.,  González-Pita A.,  Sánchez-Colina G.,
  Pérez-Penichet C.,  Waitukaitis S.,   Hidalgo R.~C.,  2014, \mn@doi
  [Geophysical Research Letters] {10.1002/2014GL059229}, 41, 3032

\bibitem[\protect\citeauthoryear{{Ambroso}, {Santore}, {Abate}  \&
  {Durian}}{{Ambroso} et~al.}{2005}]{ambroso2005a}
{Ambroso} M.~A.,  {Santore} C.~R.,  {Abate} A.~R.,   {Durian} D.~J.,  2005,
  \mn@doi [Physical Review E] {10.1103/PhysRevE.71.051305}, \href
  {http://adsabs.harvard.edu/abs/2005PhRvE..71e1305A} {71, 051305}

\bibitem[\protect\citeauthoryear{{Cadu} et~al.,}{{Cadu}
  et~al.}{2016}]{Cadu2016EGU}
{Cadu} A.,  et~al., 2016, in EGU General Assembly Conference Abstracts. p.
  16103

\bibitem[\protect\citeauthoryear{{Campins}, {Emery}, {Kelley}, {Fern{\'a}ndez},
  {Licandro}, {Delb{\'o}}, {Barucci}  \& {Dotto}}{{Campins}
  et~al.}{2009}]{Campins2009}
{Campins} H.,  {Emery} J.~P.,  {Kelley} M.,  {Fern{\'a}ndez} Y.,  {Licandro}
  J.,  {Delb{\'o}} M.,  {Barucci} A.,   {Dotto} E.,  2009, \mn@doi [Astronomy
  and Astrophysics] {10.1051/0004-6361/200912374}, \href
  {http://adsabs.harvard.edu/abs/2009A%26A...503L..17C} {503, L17}

\bibitem[\protect\citeauthoryear{Carrier, Olhoeft  \& Mendell}{Carrier
  et~al.}{1991}]{carrier1991}
Carrier W. D.~I.,  Olhoeft G.~R.,   Mendell W.,  1991, in Heiken G.,  Vaniman
  D.,   French B.,  eds, Lunar Sourcebook. Cambridge: Cambridge Univ. Press, p.
  475?594

\bibitem[\protect\citeauthoryear{{Colwell}}{{Colwell}}{2003}]{colwell03}
{Colwell} J.~E.,  2003, \mn@doi [Icarus] {10.1016/S0019-1035(03)00083-6}, \href
  {http://adsabs.harvard.edu/abs/2003Icar..164..188C} {164, 188}

\bibitem[\protect\citeauthoryear{{Colwell} \& {Taylor}}{{Colwell} \&
  {Taylor}}{1999}]{Colwell1999}
{Colwell} J.~E.,  {Taylor} M.,  1999, \mn@doi [Icarus]
  {10.1006/icar.1998.6073}, \href
  {http://adsabs.harvard.edu/abs/1999Icar..138..241C} {138, 241}

\bibitem[\protect\citeauthoryear{{Colwell}, {Brisset}, {Dove}, {Whizin},
  {Nagler}  \& {Brown}}{{Colwell} et~al.}{2015}]{Colwell2015}
{Colwell} J.,  {Brisset} J.,  {Dove} A.,  {Whizin} A.,  {Nagler} H.,   {Brown}
  N.,  2015, European Planetary Science Congress 2015, \href
  {http://adsabs.harvard.edu/abs/2015EPSC...10..766C} {10, EPSC2015}

\bibitem[\protect\citeauthoryear{{Coradini} et~al.,}{{Coradini}
  et~al.}{2011}]{coradini2011}
{Coradini} A.,  et~al., 2011, \mn@doi [Science] {10.1126/science.1204062},
  \href {http://adsabs.harvard.edu/abs/2011Sci...334..492C} {334, 492}

\bibitem[\protect\citeauthoryear{{Delbo}, {Mueller}, {Emery}, {Rozitis}  \&
  {Capria}}{{Delbo} et~al.}{2015}]{delbo2015}
{Delbo} M.,  {Mueller} M.,  {Emery} J.~P.,  {Rozitis} B.,   {Capria} M.~T.,
  2015, in {Michel} P.,  {DeMeo} F.~E.,   {Bottke} W.~F.,  eds, Asteroids IV.
  pp 107--128

\bibitem[\protect\citeauthoryear{{Fibre Verte}}{{Fibre
  Verte}}{2013}]{FibreVerte}
{Fibre Verte} 2013, Technical report, {Fiche technique reference : SQ1-25}.
{Fibre Verte}

\bibitem[\protect\citeauthoryear{Fujiwara et~al.,}{Fujiwara
  et~al.}{2006}]{fujiwara2006}
Fujiwara A.,  et~al., 2006, \mn@doi [Science] {10.1126/science.1125841}, 312,
  1330

\bibitem[\protect\citeauthoryear{{Garcia}, {Murdoch}  \& {Mimoun}}{{Garcia}
  et~al.}{2015}]{garcia15}
{Garcia} R.~F.,  {Murdoch} N.,   {Mimoun} D.,  2015, \mn@doi [Icarus]
  {10.1016/j.icarus.2015.02.014}, \href
  {http://adsabs.harvard.edu/abs/2015Icar..253..159G} {253, 159}

\bibitem[\protect\citeauthoryear{Goldman \& Umbanhowar}{Goldman \&
  Umbanhowar}{2008}]{goldman08}
Goldman D.~I.,  Umbanhowar P.,  2008, \mn@doi [Phys. Rev. E]
  {10.1103/PhysRevE.77.021308}, 77, 021308

\bibitem[\protect\citeauthoryear{Gudhe, Yalamanchili  \& Massoudi}{Gudhe
  et~al.}{1994}]{gudhe94}
Gudhe R.,  Yalamanchili R.,   Massoudi M.,  1994, \mn@doi [International
  Journal of Non-Linear Mechanics] {10.1016/0020-7462(94)90047-7}, 29, 1

\bibitem[\protect\citeauthoryear{{Gundlach} \& {Blum}}{{Gundlach} \&
  {Blum}}{2013}]{Gundlach2013}
{Gundlach} B.,  {Blum} J.,  2013, \mn@doi [Icarus]
  {10.1016/j.icarus.2012.11.039}, \href
  {http://adsabs.harvard.edu/abs/2013Icar..223..479G} {223, 479}

\bibitem[\protect\citeauthoryear{{Ho} et~al.,}{{Ho} et~al.}{2016}]{Ho2016}
{Ho} T.~M.,  et~al., 2016, in EGU General Assembly Conference Abstracts. p.
  16163

\bibitem[\protect\citeauthoryear{{Holsapple}}{{Holsapple}}{1993}]{holsapple1993}
{Holsapple} K.~A.,  1993, \mn@doi [Annual Review of Earth and Planetary
  Sciences] {10.1146/annurev.ea.21.050193.002001}, \href
  {http://adsabs.harvard.edu/abs/1993AREPS..21..333H} {21, 333}

\bibitem[\protect\citeauthoryear{Israr, Rivallant, Bouvet  \& Barrau}{Israr
  et~al.}{2014}]{israr2014}
Israr H.~A.,  Rivallant S.,  Bouvet C.,   Barrau J.-J.,  2014, Composites Part
  A: Applied Science and Manufacturing, 62, 16

\bibitem[\protect\citeauthoryear{{Jaumann} et~al.,}{{Jaumann}
  et~al.}{2012}]{jaumann2012}
{Jaumann} R.,  et~al., 2012, \mn@doi [Science] {10.1126/science.1219122}, \href
  {http://adsabs.harvard.edu/abs/2012Sci...336..687J} {336, 687}

\bibitem[\protect\citeauthoryear{{Karatekin} et~al.,}{{Karatekin}
  et~al.}{2016}]{karatekin2016}
{Karatekin} {\"O}.,  et~al., 2016, in EGU General Assembly Conference
  Abstracts. p. 17097

\bibitem[\protect\citeauthoryear{{Katsuragi} \& {Durian}}{{Katsuragi} \&
  {Durian}}{2007}]{katsuragi07}
{Katsuragi} H.,  {Durian} D.~J.,  2007, \mn@doi [Nature Physics]
  {10.1038/nphys583}, \href {http://adsabs.harvard.edu/abs/2007NatPh...3..420K}
  {3, 420}

\bibitem[\protect\citeauthoryear{Katsuragi et~al.}{Katsuragi
  et~al.}{2016}]{Katsuragi2016}
Katsuragi H.,  et~al., 2016, Physics of Soft Impact and Cratering.
Springer

\bibitem[\protect\citeauthoryear{Khaleghi, Khamis, Karray  \& Razavi}{Khaleghi
  et~al.}{2013}]{Khaleghi2013}
Khaleghi B.,  Khamis A.,  Karray F.~O.,   Razavi S.~N.,  2013, \mn@doi
  [Information Fusion] {http://dx.doi.org/10.1016/j.inffus.2011.08.001}, 14, 28

\bibitem[\protect\citeauthoryear{Kleinhans, Markies, de Vet, in~'t Veld  \&
  Postema}{Kleinhans et~al.}{2011}]{kleinhans11}
Kleinhans M.~G.,  Markies H.,  de Vet S.~J.,  in~'t Veld A.~C.,   Postema
  F.~N.,  2011, J. Geophys. Res., 116

\bibitem[\protect\citeauthoryear{{Lauretta} et~al.,}{{Lauretta}
  et~al.}{2012}]{Lauretta2012}
{Lauretta} D.~S.,  et~al., 2012, in Asteroids, Comets, Meteors 2012. p.~6291

\bibitem[\protect\citeauthoryear{McKay, Carter, Boles, Allen  \& Allton}{McKay
  et~al.}{1994}]{McKay1994}
McKay D.~S.,  Carter J.~L.,  Boles W.~W.,  Allen C.~C.,   Allton J.~H.,  1994,
  in Engineering, Construction, and Operations in Space IV. American Society of
  Civil Engineers, pp 857--866

\bibitem[\protect\citeauthoryear{Michel, Benz, Tanga  \& Richardson}{Michel
  et~al.}{2001}]{michel01}
Michel P.,  Benz W.,  Tanga P.,   Richardson D.~C.,  2001, \mn@doi [Science]
  {10.1126/science.1065189}, 294, 1696

\bibitem[\protect\citeauthoryear{Michel et~al.,}{Michel
  et~al.}{2016}]{michel2016}
Michel P.,  et~al., 2016, \mn@doi [Advances in Space Research]
  {http://dx.doi.org/10.1016/j.asr.2016.03.031}, 57, 2529

\bibitem[\protect\citeauthoryear{Murdoch, Rozitis, Green, Michel, de Lophem  \&
  Losert}{Murdoch et~al.}{2013}]{murdoch13_MNRAS}
Murdoch N.,  Rozitis B.,  Green S.~F.,  Michel P.,  de Lophem T.-L.,   Losert
  W.,  2013, \mn@doi [Monthly Notices of the Royal Astronomical Society]
  {10.1093/mnras/stt742}

\bibitem[\protect\citeauthoryear{{Murdoch}, {Sanchez}, {Schwartz}  \&
  {Miyamoto}}{{Murdoch} et~al.}{2015}]{murdochAsteroidsIV2015}
{Murdoch} N.,  {Sanchez} P.,  {Schwartz} S.,   {Miyamoto} H.,  2015, in Michel
  P.,  DeMeo F.~E.,   Bottke W.~F.,  eds, Asteroids IV. Space Science Series

\bibitem[\protect\citeauthoryear{{Murdoch} et~al.,}{{Murdoch}
  et~al.}{2016}]{Murdoch2016EGU}
{Murdoch} N.,  et~al., 2016, in EGU General Assembly Conference Abstracts. p.
  12140

\bibitem[\protect\citeauthoryear{{Nakamura} et~al.,}{{Nakamura}
  et~al.}{2008}]{Nakamura2008}
{Nakamura} A.~M.,  et~al., 2008, \mn@doi [Earth, Planets, and Space]
  {10.1186/BF03352756}, 60, 7

\bibitem[\protect\citeauthoryear{{Nakamura}, {Setoh}, {Wada}, {Yamashita}  \&
  {Sangen}}{{Nakamura} et~al.}{2013}]{Nakamura2013}
{Nakamura} A.~M.,  {Setoh} M.,  {Wada} K.,  {Yamashita} Y.,   {Sangen} K.,
  2013, \mn@doi [Icarus] {10.1016/j.icarus.2012.11.038}, \href
  {http://adsabs.harvard.edu/abs/2013Icar..223..222N} {223, 222}

\bibitem[\protect\citeauthoryear{{Newhall} \& {Durian}}{{Newhall} \&
  {Durian}}{2003}]{newhall2003}
{Newhall} K.~A.,  {Durian} D.~J.,  2003, \mn@doi [Physical Review E]
  {10.1103/PhysRevE.68.060301}, \href
  {http://adsabs.harvard.edu/abs/2003PhRvE..68f0301N} {68, 060301}

\bibitem[\protect\citeauthoryear{Pak, Van~Doorn  \& Behringer}{Pak
  et~al.}{1995}]{Pak1995}
Pak H.~K.,  Van~Doorn E.,   Behringer R.~P.,  1995, \mn@doi [Phys. Rev. Lett.]
  {10.1103/PhysRevLett.74.4643}, 74, 4643

\bibitem[\protect\citeauthoryear{Pica~Ciamarra, Lara, Lee, Goldman, Vishik  \&
  Swinney}{Pica~Ciamarra et~al.}{2004}]{Ciamarra2004}
Pica~Ciamarra M.,  Lara A.~H.,  Lee A.~T.,  Goldman D.~I.,  Vishik I.,
  Swinney H.~L.,  2004, \mn@doi [Phys. Rev. Lett.]
  {10.1103/PhysRevLett.92.194301}, 92, 194301

\bibitem[\protect\citeauthoryear{Richardson, Jr., Melosh, Greenberg  \&
  O'Brien}{Richardson et~al.}{2005}]{richardson05}
Richardson J.~E.,  Jr. Melosh H.~J.,  Greenberg R.~J.,   O'Brien D.~P.,  2005,
  \mn@doi [Icarus] {DOI: 10.1016/j.icarus.2005.07.005}, 179, 325

\bibitem[\protect\citeauthoryear{{Robinson}, {Thomas}, {Veverka}, {Murchie}  \&
  {Wilcox}}{{Robinson} et~al.}{2002}]{robinson2002}
{Robinson} M.~S.,  {Thomas} P.~C.,  {Veverka} J.,  {Murchie} S.~L.,   {Wilcox}
  B.~B.,  2002, Meteoritics and Planetary Science, \href
  {http://adsabs.harvard.edu/abs/2002M%26PS...37.1651R} {37, 1651}

\bibitem[\protect\citeauthoryear{{S{\'a}nchez} \& {Scheeres}}{{S{\'a}nchez} \&
  {Scheeres}}{2011}]{sanchez2011}
{S{\'a}nchez} P.,  {Scheeres} D.~J.,  2011, \mn@doi [The Astrophysical Journal]
  {10.1088/0004-637X/727/2/120}, \href
  {http://adsabs.harvard.edu/abs/2011ApJ...727..120S} {727, 120}

\bibitem[\protect\citeauthoryear{Scheeres et~al.,}{Scheeres
  et~al.}{2006}]{scheeres06b}
Scheeres D.,  et~al., 2006, in AIAA/AAS Astrodynamics Specialist Conference and
  Exhibit. American Institute of Aeronautics and Astronautics,
  \mn@doi{doi:10.2514/6.2006-6661}, \url
  {http://dx.doi.org/10.2514/6.2006-6661}

\bibitem[\protect\citeauthoryear{{Schwartz}, {Richardson}  \&
  {Michel}}{{Schwartz} et~al.}{2012}]{schwartz2012}
{Schwartz} S.~R.,  {Richardson} D.~C.,   {Michel} P.,  2012, \mn@doi [Granular
  Matter] {10.1007/s10035-012-0346-z}, \href
  {http://link.springer.com/article/10.1007\%2Fs10035-012-0346-z} {14, 363}

\bibitem[\protect\citeauthoryear{Scott \& Saaj}{Scott \&
  Saaj}{2009}]{scott2009}
Scott G.,  Saaj C.,  2009, in {AIAA SPACE 2009 Conference \& Exposition}. SPACE
  Conferences and Exposition

\bibitem[\protect\citeauthoryear{Seguin, Lefebvre-Lepot, Faure  \&
  Gondret}{Seguin et~al.}{2016}]{seguin2016}
Seguin A.,  Lefebvre-Lepot A.,  Faure S.,   Gondret P.,  2016, \mn@doi [Eur.
  Phys. J. E] {10.1140/epje/i2016-16063-0}, 39, 63

\bibitem[\protect\citeauthoryear{{Sullivan}, {Thomas}, {Murchie}  \&
  {Robinson}}{{Sullivan} et~al.}{2002}]{sullivan2002}
{Sullivan} R.~J.,  {Thomas} P.~C.,  {Murchie} S.~L.,   {Robinson} M.~S.,  2002,
  Asteroids III, \href {http://adsabs.harvard.edu/abs/2002aste.conf..331S} {pp
  331--350}

\bibitem[\protect\citeauthoryear{Sunday et~al.,}{Sunday
  et~al.}{2016}]{sunday2016}
Sunday C.,  et~al., 2016, \mn@doi [Review of Scientific Instruments]
  {http://dx.doi.org/10.1063/1.4961575}, 87

\bibitem[\protect\citeauthoryear{{Thomas} et~al.,}{{Thomas}
  et~al.}{2002}]{thomas02}
{Thomas} P.~C.,  et~al., 2002, \mn@doi [Icarus] {10.1006/icar.2001.6755}, \href
  {http://adsabs.harvard.edu/abs/2002Icar..155...18T} {155, 18}

\bibitem[\protect\citeauthoryear{Tsimring \& Volfson}{Tsimring \&
  Volfson}{2005}]{Tsimring2005}
Tsimring L.~S.,  Volfson D.,  2005, in Powders and Grains.

\bibitem[\protect\citeauthoryear{Tsuchiyama et~al.,}{Tsuchiyama
  et~al.}{2011}]{tsuchiyama2011}
Tsuchiyama A.,  et~al., 2011, \mn@doi [Science] {10.1126/science.1207807}, 333,
  1125

\bibitem[\protect\citeauthoryear{Tsuda, Yoshikawa, Abe, Minamino  \&
  Nakazawa}{Tsuda et~al.}{2013}]{tsuda2013}
Tsuda Y.,  Yoshikawa M.,  Abe M.,  Minamino H.,   Nakazawa S.,  2013, \mn@doi
  [Acta Astronautica] {http://dx.doi.org/10.1016/j.actaastro.2013.06.028}, 91,
  356

\bibitem[\protect\citeauthoryear{{Uehara}, {Ambroso}, {Ojha}  \&
  {Durian}}{{Uehara} et~al.}{2003}]{uehara2003}
{Uehara} J.~S.,  {Ambroso} M.~A.,  {Ojha} R.~P.,   {Durian} D.~J.,  2003,
  \mn@doi [Physical Review Letters] {10.1103/PhysRevLett.90.194301}, \href
  {http://adsabs.harvard.edu/abs/2003PhRvL..90s4301U} {90, 194301}

\bibitem[\protect\citeauthoryear{Veverka et~al.,}{Veverka
  et~al.}{2000}]{veverka2000}
Veverka J.,  et~al., 2000, \mn@doi [Science] {10.1126/science.289.5487.2088},
  289, 2088

\bibitem[\protect\citeauthoryear{Walsh, Holloway, Habdas  \& de Bruyn}{Walsh
  et~al.}{2003}]{walsh2003}
Walsh A.~M.,  Holloway K.~E.,  Habdas P.,   de Bruyn J.~R.,  2003, \mn@doi
  [Phys. Rev. Lett.] {10.1103/PhysRevLett.91.104301}, 91, 104301

\bibitem[\protect\citeauthoryear{{YEI Technology}}{{YEI
  Technology}}{2013}]{YEI}
{YEI Technology} 2013, Technical report, {YEI 3-Space Sensor Data-logging
  Technical Brief}.
{YEI Technology}

\bibitem[\protect\citeauthoryear{{de Bruyn} \& {Walsh}}{{de Bruyn} \&
  {Walsh}}{2004}]{Bruyn2004}
{de Bruyn} J.~R.,  {Walsh} A.~M.,  2004, \mn@doi [Canadian Journal of Physics]
  {10.1139/p04-025}, \href {http://adsabs.harvard.edu/abs/2004CaJPh..82..439D}
  {82, 439}

\makeatother
\end{thebibliography}

\end{document}